\documentstyle[nato,numreferences,epsf]{crckapb}
\newcommand{\bk}{{\bf k}}
\newcommand{\bx}{{\bf x}}
\begin{opening}
\title{Dynamics and transport \protect\\ near quantum-critical points}
\author{Subir Sachdev}
\institute{Department of Physics, P.O. Box 208120, \protect\\
Yale University, New Haven, CT 06520-8120, USA}
\end{opening}
\begin{document}
\runningtitle{Quantum phase transitions}
\begin{abstract}
The physics of non-zero temperature dynamics and transport near
quantum-critical points is discussed by a detailed study
of the $O(N)$-symmetric, relativistic, quantum field theory
of a $N$-component scalar field in $d$ spatial dimensions. 
A great deal of insight is gained from a simple, exact
solution of the long-time dynamics for the $N=1$ $d=1$
case: this model describes the critical point of
the Ising chain in a transverse field, and the  
dynamics in all the distinct,
limiting, physical regions of its finite temperature 
phase diagram is obtained.
The $N=3$, $d=1$ model describes insulating,
gapped, spin chain compounds: the exact,
low temperature value of the spin diffusivity is computed,
and compared with NMR experiments. 
The $N=3$, $d=2,3$ models describe
Heisenberg antiferromagnets with collinear N\'{e}el
correlations, and experimental realizations of quantum-critical
behavior in these systems are discussed.
Finally, the $N=2$, $d=2$ model describes 
the superfluid-insulator transition in lattice boson systems:
the frequency and temperature dependence of the 
the conductivity at the quantum-critical coupling is described
and implications for experiments in two-dimensional thin films
and inversion layers are noted.
\end{abstract}

{\tt 
\begin{center}
To appear in the proceedings of the \\
NATO Advanced Study Institute on \\
{\bf Dynamical properties of unconventional magnetic systems},
Geilo, Norway April 2-12, 1997,\\
edited by A. Skjeltorp and D. Sherrington,\\
Kluwer Academic, Dordrecht.\\
Report number cond-mat/9705266.
\end{center}
}

\section{Introduction}
\label{intro}

Consider a quantum system on an infinite lattice described by the Hamiltonian
${\cal H}(g)$, with $g$ a dimensionless coupling constant. For any reasonable
$g$,  all observable properties of the ground state of ${\cal H}$
 will vary smoothly as $g$ 
is varied. However, there may be special points, like $g=g_c$, where there is
a non-analyticity in some property of the ground state: we identify $g_c$
as the position of a quantum phase transition. In finite lattices, 
non-analyticities can only occur at level crossings; the possibilities in
infinite systems are richer as avoided level crossings can become sharp in the 
thermodynamic limit. In this paper, I will restrict my discussion to second
order quantum transitions, or transitions in which the 
length and time scales over which the degrees of freedom are correlated
diverge as $g$ approaches $g_c$. As I will discuss
below, any such quantum transition can be used to define a continuum quantum
field theory (CQFT): the CQFT has no intrinsic short-distance (or
ultraviolet) cutoff.
The main purpose of this paper is to 
discuss some properties 
of ${\cal H} (g)$ at
finite temperatures ($T$) in the
vicinity of $g=g_c$, in the context of some simple, but experimentally important, models. 
These studies are equivalent to a determination of the finite $T$
crossovers of the associated CQFTs. 

We begin by stating some basic concepts on the relationship between quantum critical points 
and CQFT's~\cite{bgz,zj,statphys}. 
As correlations become long range in time in the vicinity of the
critical  point, every system must be characterized by an experimentally measurable energy
scale, 
$\Delta$ which vanishes at $g=g_c$. Convenient choices are an energy gap, if one exists,
or a stiffness of an ordered phase to changes in the orientation of an order parameter.
In the models we shall consider here, $\Delta$ vanishes as a power-law as $g$
approaches $g_c$:
\begin{equation} 
\Delta \sim \bar{\Lambda} | g - g_c |^{z\nu}
\label{Gg}
\end{equation}
where $\bar{\Lambda}$ is an ultraviolet cutoff, measured in units of 
energy, $z$ is the dynamic exponent, and $\nu$ is the correlation length 
exponent~\cite{hertz,continen}.
From the perspective of a field theorist,
 the CQFT associated with the quantum
critical point is now defined by taking the limit $\\bar{\Lambda} \rightarrow \infty$ at fixed 
$\Delta$; from
(\ref{Gg}) we see that, because $z\nu > 0$, it is possible to take this limit by 
tuning the bare
coupling $g$ closer and closer to the critical point as $\bar{\Lambda}$ increases.
(A condensed matter physicist would take the complementary, but equivalent, perspective
of keeping $\bar{\Lambda}$ fixed but moving closer to criticality by lowering his probe frequency
$\omega \sim \Delta$).
Assuming the $\bar{\Lambda} \rightarrow \infty$ limit exits, the resulting CQFT then contains 
only
the energy scale
$\Delta$. At non-zero temperatures, there is a second energy scale  $k_B T$; its thermodynamic
properties will then be a {\em universal\/} function of the only dimensionless ratio
available---$\Delta/k_B T$. 

\begin{figure}
\epsfxsize=4in
\centerline{\epsffile{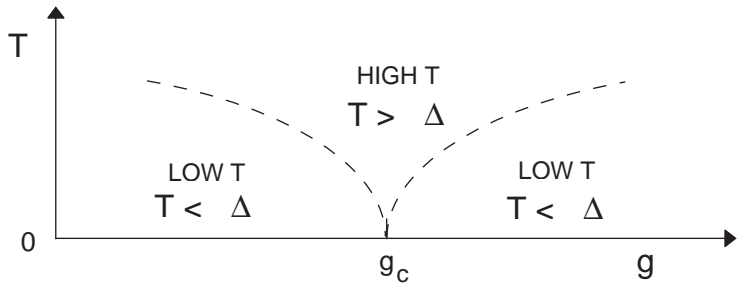}}
\caption{Schematic phase diagram as a function of the coupling constant of the quantum
Hamiltonian $g$, and the temperature $T$. The quantum critical point is at $T=0$, $g=g_c$.
The
dashed lines indicate crossovers. There may finite temperature phase transitions in either of the
two low temperature regimes. The symbol $\Delta$ represents a characteristic energy scale
of the $T=0$ theory; it vanishes at $g$ approaches 
$g_c$ according to (\protect\ref{Gg}).}
\label{fig1}
\end{figure}
Our study will find two regimes with very different physical properties, as sketched in
Fig~\ref{fig1}:
\newline
{\em (i) The low temperature region $k_B T \ll \Delta$}
\newline
There are actually two regimes of this type, one on either side of $g_c$.
Correlations in this region are similar to those of the $T=0$ ground state. The low
temperature creates a small density of excitations which can sometimes have significant
effects at very long scales.
\newline
{\em (ii) The high temperature region $\Delta \ll k_B T$}
\newline
As we are discussing universal properties of the CQFT, it is implicitly assumed that
all energy scales, including $k_B T$, are smaller than the upper cutoff $\bar{\Lambda}$ which has
been sent to infinity; so we also require that $k_B T \ll \bar{\Lambda}$.
The thermal energy,  $k_B T$, sets the scale for all physical phenomena in this region, and
the system behaves as if it's couplings are at the $g=g_c$ critical point. We shall
devote much attention to the unfamiliar and unusual properties of this region.
It is perhaps worth noting explicitly why 
the high $T$ limit of the CQFT can be non-trivial.
A conventional high $T$ expansions of the lattice model ${\cal H}$ proceeds
with the series
\begin{equation}
\mbox{Tr} e^{- {\cal H} / k_B T} = \mbox{Tr} 1 - 
\frac{1}{k_B T} \mbox{Tr} {\cal H} + \frac{1}{2(k_B T)^2} \mbox{Tr} {\cal H}^2+ \ldots
\label{highT}
\end{equation}
The successive terms in this series are well-defined and finite because of the
ultraviolet cutoffs provided by the lattice. Further, the series is
well-behaved provided $T$ is larger than all other energy scales; in particular we
need $k_B T \gg \bar{\Lambda}$. In contrast, the CQFT was defined by the limit
$\bar{\Lambda} \rightarrow \infty$ at fixed $k_B T$, $\Delta$, and, as already stated, 
the high $T$
limit of the CQFT corresponds to the intermediate temperature range $\Delta \ll k_B T \ll
\bar{\Lambda}$ of the lattice model. It is not possible to access this temperature range
by an expansion as simple as (\ref{highT}), and more sophisticated techniques,
to be discussed here, are necessary. 

All of our explicit computations will be with models described by the same CQFT.
This is the relativistic CQFT in $d$ spatial dimensions with imaginary
time ($\tau$) action
\begin{eqnarray}
{\cal S} = && \int_0^{\hbar/k_B T} d\tau \int d^d x \left\{
\frac{1}{2} \left[ ( \partial_{\tau} \phi_{\alpha} )^2 + c^2 ( \nabla_x \phi_{\alpha}
)^2 \right.\right.\nonumber \\
&&~~~~~~~~~~~~~~~~~~~~~~~~~~~~~~~\left.\left.+ ( m_{0c}^2 + (g-g_c) ) \phi_{\alpha}^2 \right] 
+ \frac{u_0}{4!} ( \phi_{\alpha}^2 )^2 \right\}.
\label{action}
\end{eqnarray}
Here $\phi_{\alpha}$ is a real scalar field, the index $\alpha = 1\ldots N$ is implicitly
summed over,
and the action has $O(N)$
symmetry. This CQFT
has a ``Lorentz'' invariance with $c$ the velocity of ``light'', and as a result
the dynamic critical exponent $z=1$.
The bare ``mass'' term has been written as $m_{0c}^2 + (g-g_c)$ so that the
$T=0$ quantum critical point is at $g=g_c$, and $u_0$ measures the strength 
of the quartic non-linearity.
The action ${\cal S}$ can also be interpreted as the Gibbs weight of a classical statistical
mechanics problem in $d+1$ dimensions (with the $\tau$ dimension of finite extent $\hbar/k_B T$);
indeed it is nothing but the standard, thoroughly-studied  $\phi^4$ theory which 
is the corner-stone of the well established theory of classical critical phenomena~\cite{bgz}.
It might then appear that we can simply carry over these results to the 
case of the quantum-critical
point, and our job is relatively straightforward. {\it This is far from being the case.}
The fundamental reason is that all dynamic experimental measurements are in real time ($t$),
and we are particularly interested in the long time limit $t \gg \hbar / k_B T$.
While, in principle, information on these long-time correlations is related by analytic
continuation to imaginary time correlations in the 
domain $0 < \tau < \hbar / k_B T$, in practice the continuation is an ill-posed problem, 
and essentially
impossible to carry out. In particular, it has been shown~\cite{sy,CSY,epsilon} 
that the operations of expansion
in $\epsilon = 3-d$ or $1/N$ (which are the only non-numeric 
tools available for analyzing ${\cal S}$),
and of analytic continuation {\em do not commute}.
It is essential that the theory of the dynamic and transport properties of ${\cal S}$
be formulated directly in real time, and here we shall review recent progress in this direction.
It is sobering to note that there remain open questions on experimentally important
observables even for the simple model ${\cal S}$, and we shall also note them here.

A central concept in our description of the dynamic
properties of ${\cal S}$ is that of the phase relaxation time, $\tau_{\varphi}$. 
This is defined as the time over which the wave-functional of the CQFT
retains phase memory. We shall find that in the regions of Fig~\ref{fig1}
\begin{eqnarray}
\tau_{\varphi} &\sim & \hbar/k_B T~~~~\mbox{in the ``High T'' region} \nonumber \\
\tau_{\varphi} &\gg & \hbar/k_B T~~~~\mbox{in the ``Low T'' regions} 
\end{eqnarray} 
These relations will be shown to apply to ${\cal S}$, but are expected to be far more general.
The missing constant in the first of these relations is a universal number which
depends upon the precise definition of $\tau_{\varphi}$.
All of the physical properties of ${\cal S}$ will show 
significant crossovers at times of order $\tau_{\varphi}$ which we shall
describe.

\section{The Ising chain in a transverse field}
\label{chap:ising}
We will begin by obtaining exact results for the dynamic scaling functions
of ${\cal S}$ for the case $d=1$, $N=1$. 
The quantum physics is more transparent in a lattice
Hamiltonian formulation, where the properties of ${\cal S}$ are expected
to be equivalent to those of the Ising chain in a transverse field 
(this may be shown in a manner similar to Section~\ref{sec:largeN}
which considers $N>1$).
The degrees of freedom of the Ising model are spins $\sigma^z_{i}$ on the
sites, $i$, of a chain. In addition to their usual exchange interaction
$J$, there is a transverse field of strength $gJ$ which is responsible
for the quantum dynamics; thus the Hamiltonian is 
\begin{equation}
H_I = -J \sum_{i} \left(g \sigma^x_i + \sigma^z_{i} \sigma^z_{i+1} \right)
\label{ising1}
\end{equation}
Like ${\cal S}$ for $N=1$, $H_I$ possesses a global $Z_2$ symmetry:
it is invariant under a unitary transformation performed by the operator
$\prod_i \sigma^x_i$ under which
\begin{equation}
\sigma^z_i \rightarrow -\sigma^z_i~~~~~~~~\sigma^x_i \rightarrow \sigma^x_i
\label{ising1a}
\end{equation}
Notice that this global Ising $Z_2$ symmetry is present in the presence of the transverse
field. A longitudinal field, coupling to $\sigma^z$ would break the $Z_2$ symmetry.
The action of this symmetry correctly indicates that the operator correspondence $\sigma^z 
\sim \phi$ maps the long distance correlators of ${\cal S}$ and $H_I$. 

\subsection{Limiting cases}
\label{sec:limiting}
We begin by examining the spectrum of $H_I$ under strong ($g \gg 1$) and weak ($g
\ll 1$) coupling~\cite{kogut}. The analysis is relatively straightforward in these limits, and
two very different physical pictures emerge. The exact solution, to be discussed later, shows
that there is a critical point exactly at
$g=1$, but that the qualitative properties of the ground states for $g>1$ ($g<1$) are very 
similar
to those for $g\gg 1$ ($g \ll 1$). One of the two limiting descriptions is therefore always
appropriate, and only the critical point
$g=1$ has genuinely different properties.

\subsubsection{Strong coupling $g \gg 1$}
Let $\left|\uparrow\right\rangle_i$ and $|\downarrow\rangle_i$ denote the eigenstates of
$\sigma^z_i$. Then $|\pm \rangle_i = (\left|\uparrow\right\rangle_i \pm 
|\downarrow\rangle_i)/\sqrt{2}$ are
the eigenstates of $\sigma^x_i$. Then at $g=\infty$ the ground state of $H_I$ is clearly
determined by  the transverse field term to be
\begin{equation}
|0\rangle = \prod_i |+\rangle_i
\label{ising1aa}
\end{equation}
The values of $\sigma^z_i$ on different sites are totally uncorrelated in this state,
and so $\langle 0 | \sigma^z_i \sigma^z_j | 0 \rangle = \delta_{ij}$. Perturbative corrections in
$1/g$ will build in correlations in $\sigma^z$ which increase in range at each order in $1/g$; 
for
$g$ large enough these correlations are expected to remain short-ranged, and the $\sigma^z$
correlator to decay exponentially with separation.
There is thus no magnetic long-range order and this state is a ``quantum paramagnet''.
Notice that this state is invariant under the $Z_2$ symmetry described above.

What about the excited states ? For $g=\infty$ these can also be listed exactly. The lowest 
excited
states are
\begin{equation}
|i\rangle = |-\rangle_i \prod_{j\neq i} |+\rangle_j,
\label{ising1ab}
\end{equation}
obtained by flipping the state on site $i$ to the other eigenstate of $\sigma^x$. All such states
are degenerate, and we will refer to them as the ``single-particle'' states. Similarly, the next
degenerate manifold of states are the two-particle states $\left|i,j\right\rangle$, obtained by
flipping the states at sites $i$ and $j$, and so on to the general $n$-particle states. To
leading order in
$1/g$, we can neglect the mixing between states between different particle number, and just study
how the degeneracy within each manifold is lifted. For the one-particle states, the exchange term
in $H_I$ leads only to the off-diagonal matrix element
\begin{equation}
\langle i | H_I | i+1 \rangle = -J
\end{equation}
which hops the `particle' between nearest neighbor sites. As in the tight-binding models of solid
state physics, the Hamiltonian is therefore diagonalized by going to the momentum space basis
\begin{equation}
|k \rangle = \frac{1}{\sqrt{N}} \sum_i e^{i k x_i} |i\rangle
\end{equation}
where $N$ is the number of sites. This eigenstate has energy (we have choosen an overall constant
in $H_I$ to make the energy of the ground state zero)
\begin{equation}
\varepsilon_k = Jg \left( 2 - (2/g) \cos ka + {\cal O}(1/g^2) \right)
\label{ising1b}
\end{equation}
where $a$ is the lattice spacing. The lowest energy one-particle state is therefore at
$\varepsilon_0 = 2g - 2J$

Now consider the two-particle states. As long as the two particles are well separated from each
other, the eigenstate is formed simply by taking the tensor product of two single particle
eigenstates. However these particles will collide, which will be described by a $S$ matrix. 
If we exclude the possibility of two-particle
bound states (which do not occur here), the total energy of the state is determined by the 
configuration where the particles are well separated, and is simply the sum of the single 
particle
energies. Thus the energy of a two-particle state with total momentum $k$ is given by
$E_k = \varepsilon_{k_1} + \varepsilon_{k_2}$ where $k= k_1 + k_2$. Notice that for a fixed $k$,
there is still an arbitrariness in the single particle momenta $k_{1,2}$ and so the total energy
$E_k$ can take a range of values. There is thus no definite energy momentum relation,
and we have instead a `two-particle continuum'. It should be clear, however, that the lowest
energy two-particle state in the infinite system (its ``threshold'') is at $2 \varepsilon_0$.
Similar considerations apply to the $n$-particle continua, which have thresholds at $n
\varepsilon_0$.

At next order in $1/g$ we have to account for the mixing between states with differing numbers of
particles. Non-zero matrix elements like
\begin{equation}
\langle 0 | H_I | i, i+1 \rangle = -J
\end{equation}
lead to a coupling between $n$ and $n+2$ particle states. It is clear that these will renormalize
the one-particle energies $\varepsilon_k$. However qualitative features of the spectrum will not
change, and we will still have renormalized one-particle states with a definite energy-momentum
relationship, and renormalized $n\geq 2$ particle continua with thresholds at $n \varepsilon_0$.

The same expansion in powers of $1/g$ can also be used to compute the two-particle $S$ 
matrix. If we consider the collision of two-particles with small momenta $p$ and $p'$,
then by conservation of energy and momentum there can only be two particles in the
final state, and the momenta of these particles remains $p$ and $p'$. An elementary
calculation to order $1/g$ shows then that
\begin{equation}
S_{p p'} = -1.
\label{smat}
\end{equation}
It can be shown that this result holds in limit of small $p$, $p'$ to all orders in 
$1/g$ for a large class of models like $H_I$ with further neighbor exchange. 
The particles under consideration, being simple spin flips, are evidently bosons which
experience a short-range repulsive potential. In $d=1$ any weak repulsive potential
appears arbitrarily strong in the limit of low velocities, and leads to the 
`unitarity limit' phase shift of $\pi$, which is responsible for (\ref{smat}).
For the particular nearest neighbor model $H_I$, it can be shown that there
is no particle production in collisions at any incoming momenta $p$, $p'$,
and that (\ref{smat}) holds to all orders in $1/g$ for all $p$, $p'$.
This will be shown and exploited in Section~\ref{sec:exact}.

\begin{figure}
\epsfxsize=3.0in
\centerline{\epsffile{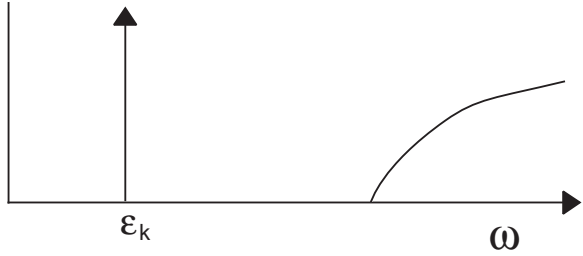}}
\caption{Schematic of the spectral density $\chi''(k,\omega)$ of $H_I$ as a function of $\omega$
at $T=0$ and a small $k$. There is a quasiparticle
delta function at $\omega = \varepsilon_k$, and a three-particle continuum at higher frequencies. 
}
\label{fig2}
\end{figure}

The spectrum described above has simple, but important, consequences for the dynamic spin
susceptibility $\chi(k, i\omega_n)$. This is defined in imaginary time as the Fourier transform
into momentum and frequency of the $\sigma^z$ correlator:
\begin{equation}
\chi (k, i\omega_n) = \int_0^{1/T} d \tau \int dx \left\langle
\sigma^z (x, i\tau) \sigma^z (0, 0) \right\rangle e^{i(kx - \omega_n \tau)}
\label{ising1c}
\end{equation}
Its spectral density $\chi''(p, \omega)$ is the imaginary part of the real frequency $\chi(k,
\omega)$, and it given by
\begin{equation}
\chi'' (k, \omega) = \pi \sum_{a} |\left\langle 0 | \sigma^z (k) | a \right\rangle |^2 
\delta(\omega - E_a)
\label{ising2}
\end{equation}
where the sum over $a$ extends over all the eigenstates of $H_I$ with energy $E_a$.
The eigenstates and energies described above allow us to simply deduce the qualitative form of
$\chi'' (p,\omega)$ which is sketched in Fig~\ref{fig2}. 
The operator $\sigma^z$ flips the state
at a single site, and so the matrix element in (\ref{ising2}) is non-zero for the single particle
states: only the state with momentum $p$ will contribute, and so there is an infinitely sharp
delta function contribution to $\chi''(k, \omega ) \sim \delta (\omega - \varepsilon_k)$. 
This delta
function is the ``quasiparticle peak'' and its co-efficient is the quasiparticle amplitude.
At $g=\infty$ this quasiparticle peak is the entire spectral density, but for smaller $g$
the quasiparticle amplitude decreases and the multiparticle states also contribute to the 
spectral
density. The mixing between the one and three particle states discussed above, means that
the next contribution to $\chi''(p,\omega)$ occurs above the 3 particle threshold $\omega > 3
\varepsilon_0$; because there are a continuum of such states, their contribution is no longer a 
delta
function, but a smooth function of omega (apart from a threshold singularity), as shown in
Fig~\ref{fig2}. 
Similarly there are continua above higher odd number particle thresholds;
only states with odd numbers of particles contribute 
because the matrix element in (\ref{ising2})
vanishes for even numbers of particles.

\subsubsection{Weak coupling $g \ll 1$}
Now the energy is dominated by the exchange term. 
There are two degenerate ground states at $g=0$ with the spins either all up or down (in
eigenstates of $\sigma^z$):
\begin{equation}
\left|\uparrow\right\rangle = \prod_i | \uparrow \rangle_i ~~~~~~~
\left|\downarrow\right\rangle = \prod_i \left|\downarrow\right\rangle_i
\end{equation}
Turning on a small $g$ will mix in a small fraction of spins of the opposite orientation, but the
degeneracy will survive as the two states are related to each other by the global $Z_2$
symmetry noted above (\ref{ising1a}). A thermodynamic system will always choose one or the other
of the states as its ground states (which may be preferred by some infinitesimal external
perturbation), and hence the $Z_2$ symmetry will be spontaneously broken. The correlations
of the magnetization $\sigma^z$ have an infinite range in either state as
\begin{equation}
\lim_{|x| \rightarrow \infty}
\left\langle \sigma^z (x, 0) \sigma^z (0,0) \right\rangle
= N_0^2 \neq 0
\label{ising2c}
\end{equation}
The quantity $N_0$ is the spontaneous magnetization, and equals $\langle \sigma^z \rangle$ in
either of the two ground states. All of the statements made in this paragraph clearly hold for
$g=0$, and will hold for some $g>0$ provided the perturbation theory in $g$ has a non-zero radius
of convergence. The exact solution of the model to be discussed later will verify that this is
indeed the case. 

The excited states can be described in terms of an elementary domain wall (or kink) excitation.
For instance the state
\begin{displaymath}
\cdots \left|\uparrow\right\rangle_i \left|\uparrow\right\rangle_{i+1}
\left|\downarrow\right\rangle_{i+2}  \left|\downarrow\right\rangle_{i+3} 
\left|\downarrow\right\rangle_{i+4}
\left|\uparrow\right\rangle_{i+5} \left|\uparrow\right\rangle_{i+6} \cdots
\end{displaymath}
has domain walls, or nearest neighbor pairs of antiparallel spins, 
between sites $i+1$, $i+2$ and sites $i+4$, $i+5$. At $g=0$ the energy of such a state is clearly
$2J \times $number of domain walls. The consequences of a small non-zero $g$ are now very
similar to those due to $1/g$ corrections in the complementary large $g$ limit: the domain walls
become ``particles'' which can hop and form momentum eigenstates with excitation energy
\begin{equation}
\varepsilon_k = J \left( 2 - 2 g \cos(ka) + {\cal O}(g^2) \right).
\label{ising2b}
\end{equation} 
The spectrum can be interpreted in terms $n$-particle scattering states, although it must be
emphasized that the interpretation of the particle is now very different from that in the large 
$g$ limit.  Again, the perturbation theory in $g$ only mixes states which differ by even numbers 
of
particles, although now the matrix element in (\ref{ising2}) is non-zero only for states $a$ with
an {\em even\/} number of particles; these assertions can easily be checked to hold in a
perturbation theory in $g$. So $\chi''(p, \omega)$ will now have a pole at $p=0$, $\omega=0^{+}$,
from the term in (\ref{ising2}) where $a =$ one of the ground states, indicating the presence of
long-range order. Further, there is now no single particle contribution, and the first finite
$\omega$ spectral density is the continuum above the two particle threshold. The absence of a
single particle delta function in this case is a very special feature of the $d=1$, $N=1$
model, and is not expected to hold in higher $d$. 

The $S$ matrix for the collision of two domain walls can now be computed in 
a perturbation theory in $g$, and as in the strong-coupling $1/g$ expansion, we find
that there is no particle production, and $S_{pp'} = -1$ to all orders in $g$.

\subsection{Exact spectrum and continuum theory}
\label{sec:exact}

The qualitative considerations of the previous section are quite useful in developing an 
intuitive
physical picture. We will now take a different route, and set up a formalism that will eventually
lead to an exact determination of many physical correlators; these results will vindicate
the approximate methods for $g>1$, $g<1$, and also provide an understanding of the novel physics
at $g=1$.

The central idea 
is the application of the Jordan-Wigner
transformation~\cite{lsm,pfeuty}. 
We map the $|+\rangle$, $|-\rangle$ states on each site to the Fock space of 
spinless fermions which can have occupation numbers $0,1$ on each site. 
The operator representation of the mapping is
\begin{eqnarray}
\sigma^x_i &=& 1 - 2 c_i^{\dagger} c_i \nonumber \\
\sigma^z_i &=& (-1/2) \prod_{j<i} \left( 1 - 2 c_j^{\dagger} c_j \right) 
(c_i + c_i^{\dagger} ) 
\label{jw1}
\end{eqnarray}
where $c_i$ is the annihilation operator for the fermion at site $i$.
Inserting (\ref{jw1}) into $H_I$ we find that the resulting Hamiltonian
is quadratic in the fermions
\begin{equation}
H_I = -J \sum_i \left(
c_i^{\dagger} c_{i+1} + c_{i+1}^{\dagger} c_i + c_i^{\dagger} c_{i+1}^{\dagger}
+ c_{i+1} c_i - 2 g c_i^{\dagger} c_i - g \right).
\end{equation}
This Hamiltonian can be diagonalized in momentum space by a Bogoliubov transformation.
The fermionic quasiparticle operators are $\gamma_k$ ($-\pi/a < k < \pi/a$)
and the diagonal form of $H_I$ is
\begin{equation}
H_I = \sum_k \varepsilon_k ( \gamma_k^{\dagger} \gamma_k - 1/2)
\end{equation}
where
\begin{equation}
\varepsilon_k = 2 J \left( 1 + g^2 - 2 g \cos (ka) \right)^{1/2}
\label{ising5}
\end{equation}
is the single particle energy. As $\varepsilon_k \geq 0$, the ground state, $|0 \rangle$, of 
$H_I$ 
has
has no $\gamma$ fermions and therefore satisfies $\gamma_k |0\rangle = 0$ for all $k$.
The excited states are created by occupying the single particle states; they can clearly be
classified by the total number of occupied states, and a $n$-particle state has the from
$\gamma_{k_1}^{\dagger} \gamma_{k_2}^{\dagger} \cdots \gamma_{k_n}^{\dagger} | 0 \rangle$,
with all the $k_i$ distinct. 

The above structure of the spectrum confirms the approximate considerations of  
Section~\ref{sec:limiting}. We have now found that the particles are in fact free fermions,
and two fermions will not scatter even when they are close to each other; alternatively they can
be considered as hard-core bosons which have an $S$ matrix which does not allow particle 
production,
and which equals $-1$ at all momenta. We shall find it much more useful to take the latter point
of view, as the bosonic particles have a simple, local, interpretation in terms of the
underlying spin excitations: for $g \gg 1$ the bosons are simply spins oriented in
the $|-\rangle$ direction, while for $g \ll 1$ they are domain walls between the two ground 
states.
The fermionic representation is useful for certain technical manipulation, but the bosonic
point of view is much more useful for making physical arguments, as we shall see below.

The excitation energy $\varepsilon_k$ in (\ref{ising5}) is non-zero and positive for all $k$ 
provided
$g\neq 1$. The energy gap, or the minimum excitation energy is always at $k=0$, and equals
$2J |1-g|$. This gap vanishes at $g=1$, and it is natural to expect that $g=g_c = 1$ is the
phase boundary between the two qualitatively different phases discussed in
Section~\ref{sec:limiting}. Precisely at $g=1$, fermions with low momenta can carry arbitrarily 
low
energy, and therefore must dominate the low temperature properties. These properties suggest that
the state at $g=1$ is critical, and there is a CQFT
which describes the critical properties in its vicinity. 
As the important excitations are near $k=0$,
we define the continuum Fermi field
\begin{equation}
\Psi(x_i) = \frac{1}{\sqrt{a}} c_i
\end{equation}
We express $H_I$ in terms of $\Psi$, and expand in spatial gradients. This
gives the fermionic Lagrangean of the CQFT believed to be equivalent to
${\cal S}$ for $d=1$, $N=1$:
\begin{equation}
{\cal L}_I = - \Psi^{\dagger} \frac{ \partial \Psi~{\dagger}}{\partial \tau}
+ \frac{c}{2} \left( \Psi^{\dagger} \frac{\partial \Psi}{\partial x}
- \Psi \frac{\partial \Psi}{\partial x} \right) + \Delta \Psi^{\dagger} \Psi
\label{ising6}
\end{equation}
The field $\Psi$ is now implicitly assumed to be a function of space and imaginary time ($\tau$).
The coupling constants in (\ref{ising6}) are given for $H_I$ by
\begin{equation}
\Delta = 2J(1-g)~~~~~~~~c = 2Ja.
\label{ising6a}
\end{equation}
However, the relations (\ref{ising6a}) are very specific to the solvable model $H_I$. For more
complicated, non-solvable, Ising models which have a similar naive continuum limit ({\em e.g.\/}
models with second neighbor exchange), the values of $\Delta$ and $c$ appearing in the continuum
quantum field theory  cannot be determined exactly. Indeed $\Delta$ and $c$ are parameters
which depend upon details of the microscopic Hamiltonian, {\em i.e.} they have a non-universal
dependence upon a coupling constant like $g$, as can be checked by a simple estimate of
perturbative fluctuation corrections due to irrelevant operators. 

Comparing the dependence of $\Delta$ on $g$ in (\ref{ising6a}) with (\ref{Gg}) we conclude
that the exponent $z\nu =1$. Further at the critical point $\Delta = 0$, the excitation
energy $\varepsilon_k \sim k$, which fixes the dynamic critical exponent $z=1$.

The continuum theory
${\cal L}_I$ can be diagonalized much like the lattice model
$H_I$, and the excitation energy now takes a ``relativistic'' form
\begin{equation}
\varepsilon_k = \left(\Delta^2 + c^2 k^2\right)^{1/2}
\end{equation}
which shows that $|\Delta|$ is the $T=0$ energy gap (we will choose the sign of $\Delta$ to be
different on the two sides of the critical value of $g$), and
$c$ is the velocity of the excitations, both measurable quantities.
The form of $\varepsilon_k$ correctly suggests that ${\cal L}_I$ is invariant under Lorentz
transformations. This can be made explicit by writing the complex Grassman field $\Psi$
in terms of two real Grassman fields, when the action becomes what is known as the field theory
of Majorana fermions of mass $\Delta/c^2$~\cite{drouffe}; we will not explicitly display this 
here.

While the continuum field $\Psi$ has no direct physical interpretation, its correlators
are easily computed at all $g$, including the critical point $g=1$. 
To gain some insight about this point, we explicitly display a certain $\Delta = 0$
two-point correlator for $T \geq 0$.
We find for imaginary time
$\tau > 0$
\begin{eqnarray}
 && \left\langle \Psi(x, i \tau) \Psi^{\dagger} (0,0) \right\rangle =
\frac{1}{2} \int_{-\infty}^{\infty} \frac{dk}{2 \pi} \frac{e^{ikx}}{e^{c |k|/T} + 1}
\left( 
e^{c |k| (1/T - \tau)} + e^{c |k| \tau} \right)
\nonumber \\
&&~~~~~~~~~~=\left( \frac{T}{4 c} \right) \left( \frac{1}{\sin(\pi T ( \tau - i x/c))}
+ \frac{1}{\sin(\pi T ( \tau + i x/c))} \right).
\label{ising6b}
\end{eqnarray}
We are now using units in which $\hbar = k_B = 1$, and will continue to do so in the remainder
of the 
paper.
At $T=0$, (\ref{ising6b}) simplifies to
\begin{equation}
\left\langle \Psi(x,\tau) \Psi^{\dagger} (0,0) \right\rangle = 
\frac{1}{4 \pi} \left( \frac{1}{c \tau - ix} + \frac{1}{c \tau + ix} \right),
\end{equation}
when we notice that the transformation 
\begin{equation}
c \tau \pm i x \rightarrow \frac{c}{\pi  T}
 \sin\left( \frac{\pi T}{c} (c \tau
\pm i x) \right)
\label{ising6c}
\end{equation}
connects the $T=0$ and $T>0$ results.
This transformation is actually a very general property of the critical point,
and connects {\it all} $T=0$ and $T>0$ correlators; it is a consequence of the conformal
invariance~\cite{cardy} of
${\cal L}_I$ at $\Delta = 0$. We will use (\ref{ising6c}) in an important way later.

Correlators of $\sigma^x$ can be constructed out of those of simple bilinears of the fermion
operators, and we will not display them explicitly. More interesting, however, are the 
correlators
of the order parameter $\sigma^z \sim \phi$.
 Computing just the equal time two-point correlator, or even
simply the value of its scaling dimension, $\mbox{dim}[\sigma^z]$, involves a rather
lengthy  and involved computation, which 
will not be discussed here. Rather, in the next section, we will quote a recently obtained
technical result, and then proceed to obtain dynamic correlators of $\sigma^z$ by
simple physical arguments.

\subsection{Finite temperature crossovers}
\label{sec:isingcrossovers}

\begin{figure}
\epsfxsize=3.6in
\centerline{\epsffile{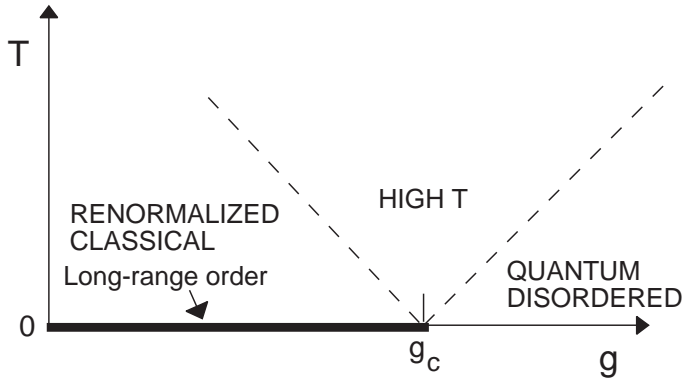}}
\caption{Finite $T$ phase diagram of the $d=1$ quantum Ising model, $H_I$, as a
function of the coupling $g$ and temperature $T$.
There is a quantum phase transition at $T=0$ $g=g_c$ with exponents $z=1$, $\nu=1$.
 Long-range order ($N_0 = \langle
\sigma_z \rangle
\neq 0$) is present only for $T=0$ and $g<g_c$. 
There is an energy gap above the ground state for all $g \neq g_c$.
We use an energy scale $\Delta \sim g_c - g$ such that the energy gap is $|\Delta|$.
The dashed lines are crossovers at $|\Delta| \sim T$. A very similar phase diagram
also applies to the $d=2$ $O(3)$ quantum rotor model discussed in 
Section~\protect\ref{sec:rotor2d}. In this case there is no energy gap
for $g<g_c$, and the spin-stiffness $\rho_s$ is used as the energy scale
to characterize the ground state. There is an energy gap $\Delta$ for
$g > g_c$. These energies vanish as $\rho_s \sim (g_c-g)^{z\nu}$,
$\Delta \sim (g-g_c)^{z\nu}$ with $z=1$, $\nu\approx 0.7$, and so the
crossover phase boundaries at $T \sim |g-g_c|^{z\nu}$ will not be linear.}
\label{fig3}
\end{figure}

The purpose of this section is to describe the dynamics of the order parameter
in the different regions of the $T>0$ phase diagram sketched in Fig~\ref{fig3};
this diagram follows from the considerations in Section~\ref{intro},
the expression for $\Delta$ in (\ref{ising6a}), and computations to be described
below.
As noted earlier, here we will quote just one technical result on the equal-time
two-point correlator of $\sigma^z$ for $T>0$ which was obtained~\cite{mccoy,barouch,ssising}
using the fermion mapping. We will then show, following the recent work
of Ref.~\cite{apy}, that unequal-time, $T>0$, correlations can be obtained by simple physical
arguments that rely on the bosonic picture of the excitations developed in
Section~\ref{sec:limiting} using the large and small $g$ expansions. 
It is worth noting that very sophisticated methods~\cite{korepbook,andre}
relying on the fermion mapping have not so far succeeded in obtaining
any explicit results for unequal time correlations for $T>0$
(there are some results~\cite{perk} on time-dependent correlators
at $T=\infty$ which are non-universal and unrelated to the CQFT of interest here),

The equal-time, long-distance result we need is~\cite{ssising}
\begin{equation}
\left\langle \sigma^z (|x| \rightarrow \infty,t=0) \sigma^z (0,0) \right\rangle
= Z T^{1/4} G_I (\Delta/T) \exp \left( - \frac{T|x|}{c} F_I (\Delta/T) \right)
\label{ising26}
\end{equation}
where $t$ is real time, $Z$ is  non-universal constant, and 
$F_I (s)$ and $G_I(s)$ are universal scaling
functions. The crucial property of (\ref{ising26}) is the prefactor
of $T^{1/4}$. As $T$ is an energy which scales as inverse time, and the dynamic
exponent $z=1$, this allows us to conclude that the scaling dimension
of $\sigma^z$ ($\phi$) is
\begin{equation}
\mbox{dim}[\sigma^z] = 1/8.
\label{dimsigma}
\end{equation}
From (\ref{ising26}), we can also define the correlation length $\xi$
which obeys
\begin{equation}
\xi^{-1} = \frac{T}{c} F_I \left( \frac{\Delta}{T} \right)
\label{ising19a}
\end{equation}
The exact, self-contained expression for the universal function $F_I$ is~\cite{ssising}
\begin{equation} 
F_I (s) = \frac{1}{\pi} \int_{0}^{\infty} dy \ln \coth \frac{(y^2 + s^2)^{1/2}}{2}  + |s| \theta 
(-s).
\label{ising22}
\end{equation}
The $s>0$ ($s<0$) portion of $F_I$ describes the ordered (disordered) side.
Despite appearances, the function $F_I (s)$ is smooth as a function of $s$ for all real $s$,
and is analytic at $s=0$. The analyticity at $s=0$ is required by the absence of any 
thermodynamic
singularity at finite $T$ for $\Delta =0$. This is a key property, which was in fact used to 
obtain 
the answer in (\ref{ising22}).
The exact expression for the function $G_I (s)$ is also known
\begin{equation}
\ln G_I (s) =    \int_{s}^{1} \frac{dy}{y} \left[ \left(
\frac{dF_I(y)}{dy} \right)^2 - \frac{1}{4} \right] + \int_{1}^{\infty} \frac{dy}{y}
\left(\frac{dF_I(y)}{dy} \right)^2 ,
\label{ising25}
\end{equation}
and its analyticity at $s=0$ follows from that of $F_I$.
For the solvable model $H_I$, we
chose the overall normalization of
$G_I$ such that
$Z=J^{-1/4}$. In general, the value of $Z$ is set by relating it to an observable,
as we will show below. Also note that $Z$ has no dependence on $\Delta$, and is therefore
non-singular at the quantum critical point.

Armed with the above knowledge, we can write down the full scaling form for
the time-dependent $\sigma^z$ correlator, which applies to the lattice model in the limits
$\bar{\Lambda} \sim J \rightarrow \infty$, $a \rightarrow 0$ at fixed $\Delta$, $c$ and $T$
\begin{equation}
\left\langle \sigma^z (x,t) \sigma^z (0,0) \right\rangle
= Z T^{1/4} \Phi_I \left( \frac{Tx}{c}, T \tau , \frac{\Delta}{T} \right)
\label{ising27}
\end{equation}
where $\Phi_I$ is a universal function which is analytic as a function of its third argument
$s=\Delta/T$ on the real $s$ axis. The result (\ref{ising26}) obviously specifies $\Phi_I$ 
for large $Tx/c$ and $t=0$.

The following subsections
will describe the unequal-time form of $\Phi_I$ in the limiting
regions of Fig~\ref{fig3}: they are associated with the limits $s \rightarrow \infty$
(renormalized classical), $s \rightarrow -\infty$ (quantum disordered),
and $s=0$ (high $T$) of $F_I$, $G_I$ which will also be noted below.

\subsubsection{Low $T$ on the ordered side, $\Delta >0$, $T \ll \Delta$} 
\label{sec:isingrc}

This is the ``renormalized classical''~\cite{CHN} region of Fig~\ref{fig3}, and the reasons for
this name will become clear below.

Assuming that it is valid to interchange the limits $T \rightarrow 0$ and $x \rightarrow \infty$
in (\ref{ising26}), we can use the limiting values $F_I (\infty) = 0$, $G_I (s \rightarrow 
\infty) =
s^{1/4}$ to deduce that (recall (\ref{ising2c})):
\begin{equation}
N_0^2 \equiv \lim_{|x| \rightarrow \infty}
\left\langle \sigma^z (x, 0) \sigma^z (0,0) \right\rangle
= Z \Delta^{1/4}~~~~~\mbox{at $T=0$.}
\label{ising27a}
\end{equation}
Thus, as claimed earlier, there is long-range order in the $g< 1$ ground state of $H_I$, with
the order parameter $N_0 = \langle \sigma^z \rangle = Z^{1/2} \Delta^{1/8}$ (this relates the 
value
of $Z$ to a physical observable). For small $T \ll \Delta$, we obtain from
the large
$s$ behavior of
$F_I (s)$ (see (\ref{ising22})) that
\begin{equation}
\left\langle \sigma^z (x, 0) \sigma^z (0,0) \right\rangle
= N_0^2 e^{-|x|/\xi_c}~~~\mbox{large $|x|$},
\label{ising28}
\end{equation}
where the correlation length
\begin{equation}
\xi_c^{-1} = \left( \frac{2 \Delta T}{\pi c^2} \right)^{1/2} e^{-\Delta/T}.
\label{ising28a}
\end{equation}
is finite at any non-zero $T$, showing that long-range order is present only
precisely at $T=0$.
We have put a subscript $c$ on the correlation length to emphasize that the system is expected to
behave {\em classically\/} in this low temperature region~\cite{ssising}. The excitations above
the ground states consists of particles (the kinks and anti-kinks of Section~\ref{sec:limiting})
whose mean separation ($\sim \xi_c$) is much larger than their de Broglie wavelengths
($\sim (c^2/\Delta T)^{1/2}$, as the mass of these particles = $\Delta/c^2$), which is precisely
the canonical condition for the applicability of classical physics. 
It is also reassuring to note that (\ref{ising28}) is precisely the form of equal-time 
correlations
in the classical Ising model at low $T$. The prefactor $N_0^2$ is the true ground state
magnetization including the effects of quantum fluctuations, and this is the reason for
the adjective ``renormalized'' in the name for this region.

We now show, following Ref~\cite{apy}, that it is possible to give a simple physical 
interpretation
for the precise value of $\xi_c$ in (\ref{ising28a}). The energy of a domain wall with a small
momentum $k$ is $\Delta + c^2 k^2 / 2 \Delta$, and therefore classical Boltzmann statistics
tells us that their density, $\rho$, is
\begin{equation}
\rho = \int \frac{dk}{2 \pi} e^{-(\Delta + c^2 k^2 /2 \Delta)/T} = 
\left( \frac{T\Delta}{2 \pi c^2} \right)^{1/2} e^{-\Delta/T}.
\label{eq:density}
\end{equation}
Comparing with (\ref{ising28a}), we see that $\xi_c = 1/ 2 \rho$. We now show that this 
result follows immediately if we assume that the domain walls are classical {\em point} 
particles,
which are distributed independently with a density $\rho$. Consider a system of size
$L \gg |x|$, and let it contain $M$ thermally excited particles; then $\rho = M/L$.
Let $q$ be the probability that a given particle will be between $0$ and $x$.
Clearly, $q = |x|/L$. The probability that a given set of $j$ particles are the only
ones between 0 and $x$ is then $q^j (1-q)^{M-j}$: as each particle reverses the orientation
of the ground state, in this case $\sigma^z (x,0) \sigma^z (0,0) = N_0^2 (-1)^j$.
Summing over all possibilities we have
\begin{eqnarray}
\langle \sigma^z (x,0) \sigma^z (0,0) \rangle 
&=& N_0^2 \sum_{j=0}^M (-1)^k q^j (1-q)^{M-j} \frac{M!}{j! (M-j)!} \nonumber \\
&=& N_0^2 ( 1 - 2 q)^N \approx N_0^2 e^{-2 q M} = N_0^2 e^{-2 \rho |x|},
\label{apy1}
\end{eqnarray}
thus establishing the desired result.

\begin{figure}
\epsfxsize=3.2in
\centerline{\epsffile{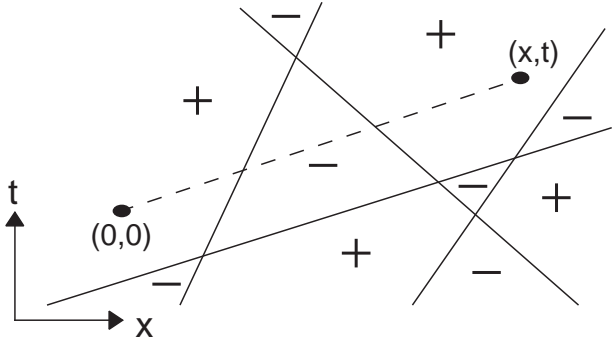}}
\caption{A typical semiclassical contribution to the double time path integral
for $\langle \sigma^z (x,t) \sigma^z (0,0) \rangle$.  
Full lines are thermally excited particles which
propagate forward and 
backward in time.  The $\pm$ signs are significant only for $g<1$
and denote the orientation of the order parameter. For $g>1$, the dashed line
is a particle propagating only forward in time from $(0,0)$ to $(x,t)$.}
\label{fig4}
\end{figure}

This classical picture can also be extended to compute unequal time correlations~\cite{apy}.
Let us explicitly consider the form of the correlation
function needed
\begin{equation}
\langle \sigma^z (x,t) \sigma^z (0,0) \rangle = 
\mbox{Tr} \left ( e^{-H_I /T} e^{i H_I t} \sigma^z_i 
e^{-i H_I t} \sigma^z_0 \right)/Z ,
\label{keldysh}
\end{equation}
where $Z = \mbox{Tr} e^{-H_I /T}$. 
We evaluate it in
the double-time (`Keldysh') path integral formalism~\cite{keldysh}. The path
integral is over a set of trajectories moving forward in time, representing the
operator $e^{-iH_I t}$,
and a second set moving backwards in time, corresponding
to the action of $e^{iH_I t}$.  In the semiclassical limit, stationary phase is
achieved when the backward paths are simply the time reverse of the
forward paths, and both sets are the classical trajectories. An example of a
set of paths is shown in Fig~\ref{fig4}.
Now observe that
({\em i\/})
the classical trajectories remain straight lines
across collisions because the momenta before and after the
collision are the same in one dimension; 
({\em ii\/}) for each collision, the amplitude for the
path acquires a phase $S_{pp'}$ along the forward path and its complex
conjugate along the backward path: the net factor for the collision is
therefore $|S_{pp'}|^2 = 1$.  These two facts imply that the trajectories are
simply independently distributed straight lines, placed with a uniform
density $\rho$ along the $x$ axis, with an inverse slope
\begin{equation}
v_k \equiv \frac{d \varepsilon_k}{dk},
\end{equation} 
and with their momenta chosen 
with the Boltzmann probability density $e^{-\varepsilon_k /T}/\rho$
(Fig~\ref{fig4}).

Computing dynamic correlators is now an exercise in classical probabilities. 
The value of 
$\sigma^z (0,0) \sigma^z (x,t)$ 
is the square of 
the magnetization renormalized by quantum fluctuations ($N_0^2$),
times $(-1)$ if
the number of
trajectories intersecting the dashed line in Fig.~\ref{fig4} is odd.
This precisely the computation carried out above for the equal
time correlations, but now it is easy to see from Fig~\ref{fig4} that
for particles with velocity $v_k$
\begin{equation}
q = \frac{|x - v_k t|}{L}
\end{equation}
Averaging over velocities, and then
proceeding as in (\ref{apy1}) we obtain one of our central results~\cite{apy,maki}
\begin{equation}
\langle \sigma^z (x,t) \sigma^z (0,0) \rangle = N_0^2
\exp\left( - \int \frac{dk}{\pi}e^{-\varepsilon_k/T}
\left| x - v_k t \right| \right).
\label{rxt}
\end{equation}
Notice that the equal-time correlator is
$N_0^2 e^{-|x|/\xi_c}$ as before, and the equal-space correlator decays
exponentially in time as
$N_0^2 e^{-|t|/\tau_c}$, but the general behavior is
more complicated. The classical relaxation time $\tau_c$ is a central
quantity; remarkably, we find from (\ref{rxt}) that the correlation time, $\tau_c$, is
independent of the functional form of $\varepsilon_k$ and depends only on the gap:
\begin{equation}
\tau_c = \frac{\pi\hbar}{2k_B T} e^{\Delta /k_B T},
\label{tauphirc}
\end{equation}
where we have momentarily inserted the fundamental constants $\hbar$, $k_B$ to emphasize
the universality of the prefactor.
As $\tau_c$ is also the time over which the phase coherence of the 
ground state is lost, we may identify the phase relaxation time of Section~\ref{intro}
as $\tau_{\varphi} = \tau_c$.

In the limit $T \ll \Delta$ we are now able to completely specify the form of the scaling
function $\Phi_I$ in (\ref{ising27}). We find that $\Phi_I$ collapses into a
{\it reduced scaling function} $\Phi_{Ic}$ which is determined by classical thermal
physics. In particular we have
\begin{equation}
\left\langle \sigma^z (x, t) \sigma^z (0,0) \right\rangle
= N_0^2 \Phi_{Ic} \left ( \frac{x}{\xi_c} , \frac{t}{\tau_c} \right),
\label{apy2}
\end{equation}
with
\begin{equation}
\ln \Phi_{Ic} (\bar{x}, \bar{t}) = - \bar{x}~\mbox{erf}
\left ( {\bar{x} \over\bar{t} \sqrt{\pi} }\right) -
 \bar{t}e^{-\bar{x}^2/(\pi \bar{t}^2)} .
\end{equation}
Notice that the characteristic time $\tau_c$ and length $\xi_c$ both
diverge as $\sim e^{\Delta /T}$, and so we can define an effective classical
dynamic exponent $z_c = 1$ (there is no fundamental reason why $z_c$ and $z$
should have the same value).
Classical scaling forms like (\ref{apy2})
have been discussed earlier~\cite{ssising}, but
it was incorrectly conjectured that the scaling functions would be those of the
Glauber model~\cite{glauber} (which has $z_c = 2$); Glauber dynamics does not conserve total
energy and momentum, and these conservation laws have played a crucial role in
the kinematic constraints on particle collisions.

All of the results above have been compared with exact 
numerical computations and the agreement is essentially perfect~\cite{apy}, giving us confidence
in the physical approach used to understand dynamic properties at $T>0$.

\subsubsection{Low $T$ on the disordered side, $\Delta < 0$, $T \ll |\Delta|$} 
\label{sec:isingqd}

This is the ``quantum disordered'' region of Fig~\ref{fig3}.

Now we need to take the $s \rightarrow -\infty$ limit of the functions $F_I (s)$, 
$G_I (s)$; from these limits we find
\begin{equation}
\left\langle \sigma^z (x, 0) \sigma^z (0,0) \right\rangle
= \frac{Z T}{|\Delta|^{3/4}} e^{- |r|/\xi}~,~\mbox{$|x| \rightarrow \infty$ at fixed $0< T \ll
|\Delta|$},
\end{equation}
with the correlation length $\xi$ given by
\begin{equation}
\xi^{-1} = \frac{|\Delta|}{c} + \left( \frac{2 |\Delta| T}{\pi c^2} \right)^{1/2} e^{-|\Delta|/T}
\end{equation}
So correlations decay exponentially on a scale $\sim c/|\Delta|$, and there is no long-range 
order.
The equal time correlations at $T=0$ behave in a similar manner, although the limits
$T \rightarrow 0$ and $|x| \rightarrow \infty$ do not commute for the prefactor of the 
exponential
decay. The $T=0$ result is determined by the form-factor expansion technique~\cite{formfac},
which we will not discuss here; it yields~\cite{babelon,andre}
\begin{equation}
\left\langle \sigma^z (x, 0) \sigma^z (0,0) \right\rangle
= Z |\Delta|^{1/4} \left( \frac{c} {2 \pi |\Delta| |x|} \right)^{1/2} e^{- \Delta
|x|/c}~,~\mbox{$|x|
\rightarrow
\infty$ at $T=0$}.
\label{ising29}
\end{equation}

The form factor expansion also yields the $T=0$ dynamic susceptibility (defined in
(\ref{ising1c})). The leading term is in fact precisely the quasiparticle pole at energy
$\varepsilon_k = (c^2 k^2 +
\Delta^2)^{1/2}$ that was argued to exist in this phase in Section~\ref{sec:limiting}. We have
\begin{equation}
\chi (k, \omega ) = \frac{2c Z |\Delta|^{1/4}}{c^2 k^2 + \Delta^2 - (\omega+i\delta)^2}
+ \ldots ~~,~\mbox{$T=0$}
\label{ising30}
\end{equation}
where $\delta$ is a positive infinitesimal. 
The imaginary part of this gives the delta function sketched in Fig~\ref{fig2}; the continuum of
excitations above the three particle threshold come from higher order terms in the form factor
expansion, and are  represented by the ellipsis in (\ref{ising30}). It can now be checked that
the Fourier transform of (\ref{ising30}) yields the leading term (\ref{ising29}) in the equal 
time
correlation function.

The result (\ref{ising30}) shows that the {\em quasiparticle residue\/} is $2 Z \Delta^{1/4}$
(this is another relation between $Z$ and a physical measurable, and along with
(\ref{ising27a}), it implies a relationship between the values of the residue and $N_0$ as we
approach the critical point from either side). The residue vanishes at the critical point $\Delta
=0$, where the quasiparticle picture breaks down, and we will have a completely different 
structure
of excitations.

The above is an essentially complete description of the correlations and excitations of the 
quantum paramagnetic ground state. 
We now turn to the dynamic properties at $T > 0 $. 
At nonzero $T$, there will be a small density of quasi-particle
excitations which will behave classically for the same reasons as in Section~\ref{sec:isingrc}:
their mean spacing is much larger than their de Broglie wavelength. 
The collisions of these thermally excited quasi-particles will lead to a broadening of the
delta function pole in (\ref{ising30}): the form of this broadening can be computed
exactly in the limit $T \ll |\Delta|$ using a semiclassical approach similar to that
employed for the ordered side~\cite{apy}. 
The argument again employs a semiclassical path-integral approach to evaluating the
correlator in (\ref{keldysh}). The key observation now is that we may consider the
operator $\sigma^z$ to be given by
\begin{equation}
\sigma^z (x,t) = 2 Z |\Delta|^{1/4} ( \psi(x,t) + \psi^{\dagger} (x,t)) + \ldots
\end{equation}
where $\psi^{\dagger}$ is the operator which creates a single particle excitation from the
ground state, and the ellipsis represent multi-particle creation/annihilation terms which 
are subdominant in the long time limit. This representation may also be understood from
the $g \gg 1$ picture discussed earlier, in which the single-particle excitations where
$|-\rangle$ spins: the $\sigma^z$ operator flips spins between the $\pm x$ directions,
and therefore creates and annihilates quasiparticles.

The computation of the nonzero $T$ relaxation is best done in real space and time,
so let us first write down the $T=0$ correlations in this representation.
We define $K(x,t)$ to be $T=0$ correlator of the order parameter:
\begin{eqnarray}
K(x,t) &\equiv& \langle \sigma^z (x,t) \sigma^z (0,0) \rangle_{T=0} \nonumber \\
&=& \int \frac{dk}{2 \pi} \frac{ c Z|\Delta|^{1/4}}{\varepsilon_k} e^{ikx - \varepsilon_k t} 
\nonumber 
\\
&=& \frac{Z |\Delta|^{1/4}}{\pi} K_0 ( |\Delta| (x^2 - c^2 t^2)^{1/2} /c)
\label{defk}
\end{eqnarray}
where $K_0$ is the modified Bessel function. This result is just the Fourier transform
of (\ref{ising30}. Note that for $t >
|x|/c$, the Bessel function has imaginary argument and is therefore complex and oscillatory.

Now we consider the $T\neq 0$ evaluation of (\ref{keldysh}) 
in the semiclassical path-integral approach~\cite{apy}. 
A typical set of paths 
contributing to the Keldysh path integral is still given by
Fig~\ref{fig4}, but its physical interpretation is now very different. The
dashed line now represents the trajectory of a particle created at $(0,0)$ and
annihilated at $(x,t)$, and $\pm$ signs in the domains should be ignored. In
the absence of any other particles this dashed line would contribute $K(x,t)$
to $\langle \sigma^z (x,t) \sigma^z (0,0) \rangle$. 
The scattering off the background particles (the
full lines in Fig~\ref{fig4}) introduces factors of the $S$-matrix element $S_{pp'}$;
as the dashed line only propagates forward in time, the
$S$-matrix elements for collisions between the dashed and full lines (and {\em
only} these) are not neutralized by a complex conjugate partner. Using the low
momentum value $S_{pp'} = -1$, we see that the contribution 
to $\langle \sigma^z (x,t) \sigma^z (0,0) \rangle$ 
equals $(-1)^{n_{\ell}} K(x,t)$ where $n_{\ell}$  is
the number of full lines intersecting the dashed line. The $(-1)^{n_{\ell}}$ is
precisely the term that appeared in the renormalized-classical, although for very
different reasons. We can carry out the averaging over all
trajectories as before, and obtain one of our 
main results~\cite{apy}
\begin{equation}
\langle \sigma^z (x,t) \sigma^z (0,0) \rangle = K(x,t) 
\exp\left( - \int \frac{dk}{\pi}e^{-\varepsilon_k/T}
\left| x - v_k t \right| \right) ,
\label{main_res}
\end{equation}
where $K(x,t)$ is given by Eq.~(\ref{defk}).

The result (\ref{main_res}) 
clearly displays the separation in scales at which quantum and
thermal effects act. Quantum fluctuations determine the oscillatory, complex
function $K(x,t)$, which gives the
$T=0$ value of the correlator.
Exponential relaxation of spin correlations occurs at
longer scales $\sim \xi_c, \tau_c$, and is controlled by the classical motion of
particles and a purely real relaxation function. This relaxation leads to a
broadening of the quasi-particle pole with widths of order $\xi_c^{-1}$,
$\tau_c^{-1}$ in momentum and energy space. Note that the relaxation function
is identical to that obtained in the renormalized classical region, and so
the expressions for $\xi_c$ and $\tau_c$ are the same as before,
upto the mapping $\Delta \rightarrow |\Delta|$. 
We can consider the presence of a quasi-particle pole in the response
function as a consequence of quantum coherence in the ground state,
and so the phase relaxation time, $\tau_{\varphi}$, over which this 
coherence is lost may be identified with $\tau_c$; we have therefore
\begin{equation}
\tau_{\varphi} = \frac{\pi\hbar}{2k_B T} e^{|\Delta| /k_B T},
\label{tauphiqd}
\end{equation}

\begin{figure}[t]
\epsfxsize=3.7in
\centerline{\epsffile{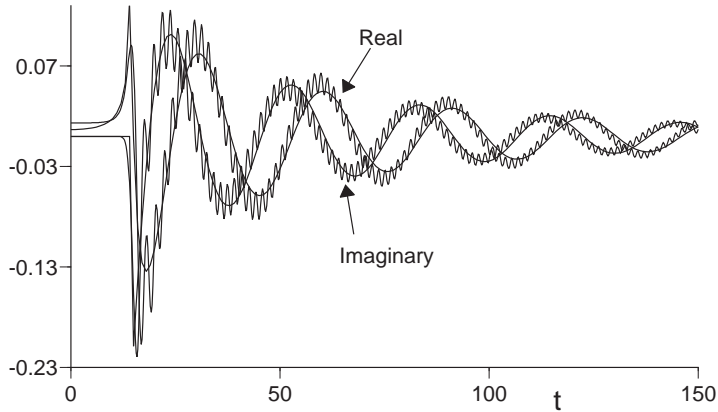}}
\caption{Theoretical and numerical results from Ref.~\protect\cite{apy} for the correlator
$\left\langle \sigma^z (x,t) \sigma^z (0,0) \right\rangle$
of $H_I$ at $x=30$ with $J=1$, $g=1.1$ (therefore $\Delta = -0.2$), $T=0.1$
and so the system is in the quantum-disordered region of Fig~\protect\ref{fig3}. 
The numerical data was
obtained for a lattice size $L=512$ with free boundary conditions;
it has a ``ringing'' at high frequency due to the lattice cutoff.
The theoretical prediction is from the continuum theory 
prediction in Eq.~(\protect\ref{main_res}) and is represented by the smoother curve.
The envelope of the numerical curve fits
the theoretical prediction
well.}
\label{fig5}
\end{figure}

In Fig.~\ref{fig5} we
compare the predictions of Eq.~(\ref{main_res}) with numerical results on a
lattice of size $L=512$. 
The theoretical curve was determined
from the continuum expression for $K(x, t)$, but 
the full lattice form for $\varepsilon_k$ was used.
The theory agrees well with the numerics; some differences are
visible for small $x$, outside the light cone, but this is outside the
domain of validity of (\ref{main_res}). 

\subsubsection{High $T$, $T \gg |\Delta|$}  
\label{sec:isingqc}

Right at the critical point, $\Delta =0$, this regime extends all the way down to $T=0$.
We begin by writing the $T=0$ equal-time correlator of the
continuum theory; from the scaling dimension of $\sigma^z$ in (\ref{dimsigma}),
this must have the form
\begin{equation}
\left\langle\sigma^z (x, 0) \sigma^z (0,0) \right\rangle
\sim
 \frac{1}{(|x|/c)^{1/4}}~~~~~~\mbox{at $T=0$, $\Delta =0$},
\label{ising31}
\end{equation}
We will now fix the prefactor in (\ref{ising31}) using our earlier results.
The key ingredient is our knowledge that the underlying continuum model ${\cal L}_I$ is
is relativistically and conformally invariant at the
$T=0$ critical point. Further we assume that the mapping
(\ref{ising6c}) between correlators at $T\neq 0$ from those at $T=0$ holds also for the two
point correlator of $\sigma^z$
Then from (\ref{ising31}) we must have at $T \neq 0$, but $\Delta =0$
\begin{equation}
\left\langle\sigma^z (x, i\tau) \sigma^z (0,0) \right\rangle
\sim T^{1/4}
 \frac{1}{
\left[ \sin(\pi T ( \tau - i x/c))\sin(\pi T ( \tau + i
x/c)) \right]^{1/8}}.
\label{ising32}
\end{equation}
Notice that we are now working in imaginary time, as it is slightly more convenient
for our purposes; the real time result can in this case be obtain by analytic continuation,
as we have the {\em exact} functional form for $0 \leq \tau \leq 1/T$.
Let us now use this result in the equal-time case in the regime $xT/c \gg 1$.
Precise results for this regime where quoted earlier in (\ref{ising26}), where using
the values $F_I (0)
= \pi/4$ (from evaluation of (\ref{ising22})) and $G_I (0) = 0.858714569\ldots$ we have
\begin{equation}
\left\langle \sigma^z (|x|\rightarrow \infty,\tau=0) \sigma^z (0,0) \right\rangle
= Z T^{1/4} G_I (0) \exp \left( - \frac{\pi T|x|}{4 c}\right)~~
\mbox{at $\Delta=0$}
\label{ising33}
\end{equation}
Finally, comparing with (\ref{ising32}) we obtain, for $\Delta =0$
\begin{equation}
\left\langle\sigma^z (x, i\tau) \sigma^z (0,0) \right\rangle
= Z T^{1/4}
 \frac{2^{-1/8} G_I (0)}{
\left[ \sin(\pi T ( \tau - i x/c))\sin(\pi T ( \tau + i
x/c)) \right]^{1/8}}.
\label{ising33a}
\end{equation}
As expected, this result is of the scaling form (\ref{ising27}), and indeed completely
determines the function $\Phi_I$ for the case where its last argument is zero.

\begin{figure}[t]
\epsfxsize=3.5in
\centerline{\epsffile{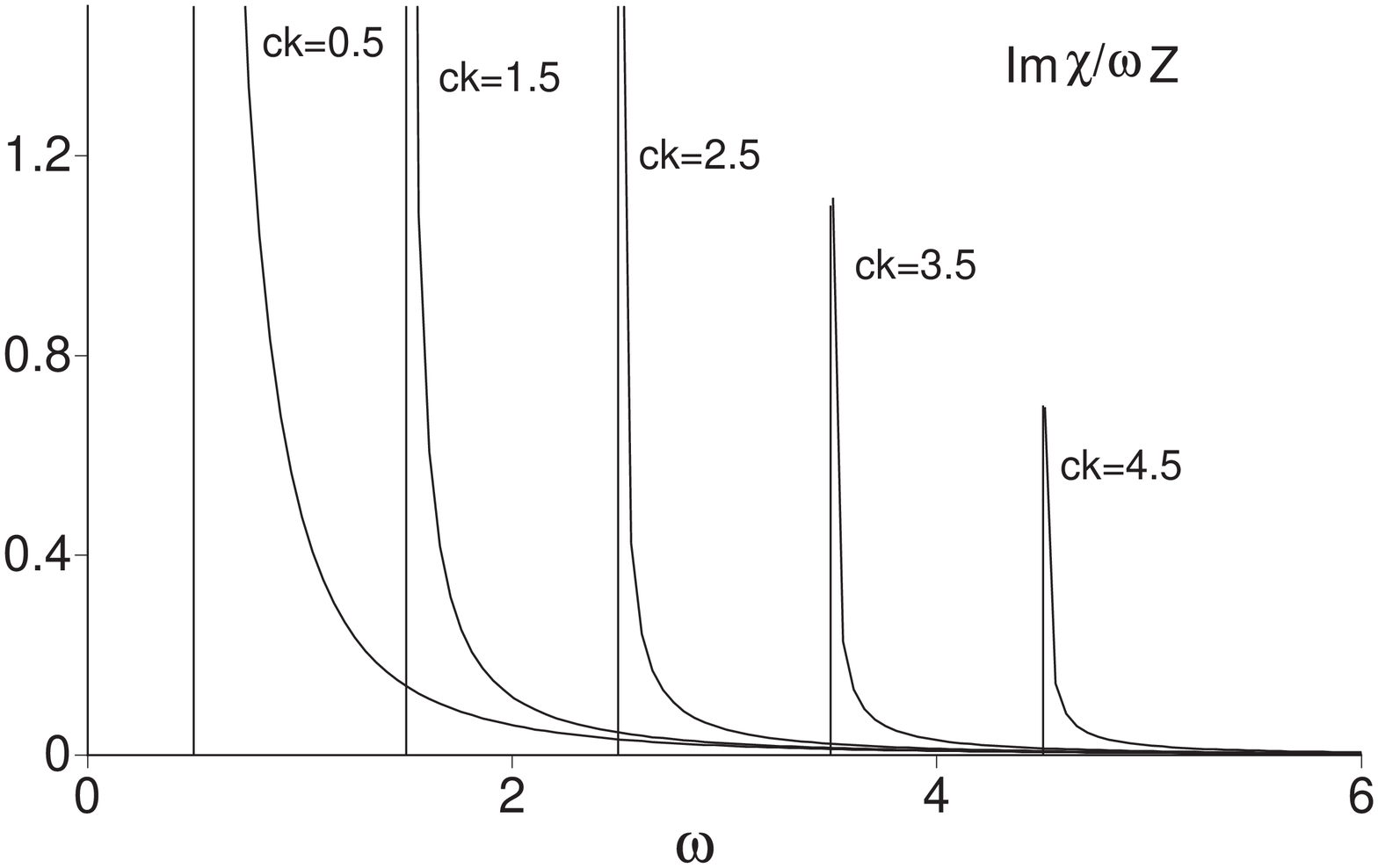}}
\caption{Spectral density, $\mbox{Im}
\chi (k,\omega )/\omega Z$, of $H_I$
at its critical point $g=1$ ($\Delta=0$) at $T=0$, as a function of frequency
$\omega$,
for a set of values of $k$.}
\label{fig6}
\end{figure}

Now let us turn to a physical interpretation of the main result (\ref{ising33a}).
Consider first the case $T=0$. By a Fourier transformation of the $T=0$ limit
of (\ref{ising33a}) we obtain the dynamic susceptibility
\begin{equation}
\chi(k, \omega) = Z ( 4 \pi)^{3/4} G_I (0) \frac{\Gamma (7/8)}{\Gamma (1/8)}
\frac{c}{(c^2 k^2 - (\omega+ i \delta)^2)^{7/8}}~~,~~\mbox{$T=0$, $\Delta=0$}
\label{ising34}
\end{equation}
with $\delta$ a positive infinitesimal.
We plot $\mbox{Im}
\chi (k,\omega )/\omega$ in  Fig~\ref{fig6}. 
Notice that there are no delta functions in the
spectral density like there were in the quantum disordered phase (Fig~\ref{fig2}), indicating the
absence of any well-defined quasiparticles. Instead, we have  a critical continuum of
excitations. However, the presence of sharp thresholds and singularities indicates
that there is still perfect phase coherence, as there must be in the ground state.

\begin{figure}[t]
\epsfxsize=3.5in
\centerline{\epsffile{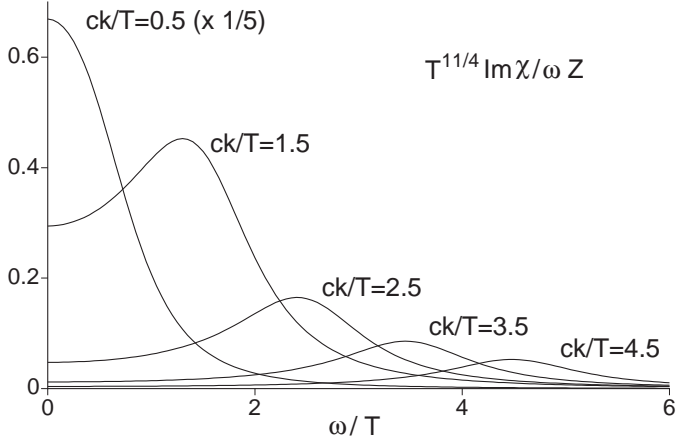}}
\caption{The same observable as in Fig~\protect\ref{fig6}, $T^{11/4} \mbox{Im}
\chi (k,\omega )/\omega Z$, but for $T \neq 0$. This is the leading result for
$\mbox{Im} \chi$ for $T \gg |\Delta|$ {\em i.e.} in the high $T$ region of 
Fig~\protect\ref{fig3}.
All quantities are scaled appropriately with powers of
$T$, and the absolute numerical values of both axes are meaningful.}
\label{fig7}
\end{figure}

Now let us turn to non-zero $T$. 
We Fourier transform (\ref{ising33a}) to obtain
$\chi(k,
i\omega_n)$ at the Matsubara frequencies $\omega_n$ and then analytically continue to real
frequencies (there are some interesting subtleties in the Fourier transform to $\chi (k,
i\omega_n)$ and its analytic structure in the complex $\omega$ plane, which are discussed
elsewhere\cite{sss}). This gives us the leading result for $\chi (k, \omega)$ in 
the high $T$ region
\begin{equation}
\chi (k , \omega ) = \frac{Zc}{T^{7/4}} \frac{G_I (0)}{4 \pi}
\frac{\Gamma(7/8)}{\Gamma(1/8)}
\frac{\displaystyle \Gamma \left( \frac{1}{16} + i \frac{\omega + ck}{4 \pi T} \right)
\Gamma \left( \frac{1}{16} + i \frac{\omega - ck}{4 \pi T} \right)}
{\displaystyle \Gamma \left( \frac{15}{16} + i \frac{\omega + ck}{4 \pi T} \right)
\Gamma \left( \frac{15}{16} + i \frac{\omega - ck}{4 \pi T} \right)}.
\label{ising35}
\end{equation}
We show a plot of $\mbox{Im} \chi /\omega$ in Fig~\ref{fig7}.
This result is 
the finite $T$ version of Fig~\ref{fig6}. Notice that the sharp features
of Fig~\ref{fig6} have been smoothed out on the scale $T$, and there is non-zero absorption at
all frequencies. For $\omega, k \gg T$ there is a
well-defined peak in $\mbox{Im} \chi /\omega$ (Fig~\ref{fig7}) rather like the $T=0$
critical behavior of Fig~\ref{fig6}. However, for $\omega, k \ll T$ we cross-over to
the {\bf quantum relaxational} regime~\cite{CSY} and the spectral density $\mbox{Im} \chi
/\omega$ is similar to (but not identical) a Lorentzian around $\omega = 0$. This relaxational
behavior can be characterized by a relaxation rate $\Gamma_R$ defined as
\begin{equation}
\Gamma_R^{-1} = -i \left. 
\frac{\partial \ln \chi (0, \omega)}{\partial \omega} \right|_{\omega =
0}
\label{ising36}
\end{equation}
(this is motivated by the phenomenological relaxational form $\chi(0, \omega) = 
\chi_0 / (1 - i \omega / \Gamma_R + {\cal O}(\omega^2))$).
The frequency scale on which the sharp features of Fig~\ref{fig6} have been smoothed out
is also $\Gamma_R$, and so we may now identify the phase coherence time $\tau_{\varphi}
= 1/\Gamma_R$.
From (\ref{ising35}) we determine:
\begin{equation}
\frac{1}{\tau_{\varphi}} = \Gamma_R = \left( 2 \tan \frac{\pi}{16} \right) \frac{k_B T}{\hbar},
\label{ising37}
\end{equation}
where we have returned to physical units. 
At the scale of the characteristic rate $\Gamma_R$,
the dynamics of the system involves intrinsic quantum effects (responsible for the
non-Lorentzian lineshape) which cannot be neglected; description by an effective classical
model (as was appropriate in both the renormalized classical and quantum disordered regions
of Fig~\ref{fig3})
would require that
$\Gamma_R
\ll k_B T /
\hbar$, which is thus not satisfied in the high $T$ region of Fig~\ref{fig3}. 
Alternatively stated, the mean spacing between the thermally excited particles is now
of order their de-Broglie wavelength, and so the classical thermal and quantum 
fluctuations must be treated at an equal footing.

The ease with
which our expressions for $\tau_{\varphi}$ in (\ref{tauphirc},\ref{tauphiqd},\ref{ising37})
have been obtained bely their remarkable nature. Notice that we are working
in a closed Hamiltonian system, evolving unitarily in time with the operator
$e^{-i H_I t/\hbar}$, from an initial density matrix given by the Gibbs ensemble at a
temperature
$T$. Yet, we have obtained relaxational behavior at low frequencies, and determined
exact values for a dissipation constant. 
In contrast, in the theory of dynamics near classical critical
points~\cite{halphoh}, the relaxation dynamics is simply postulated 
in a rather {\em ad hoc} manner, and the relaxational constants
are treated as phenomenological parameters to be determined by comparison
with experiments.

\subsubsection{Summary} 
\begin{figure}[t]
\epsfxsize=3.3in
\centerline{\epsffile{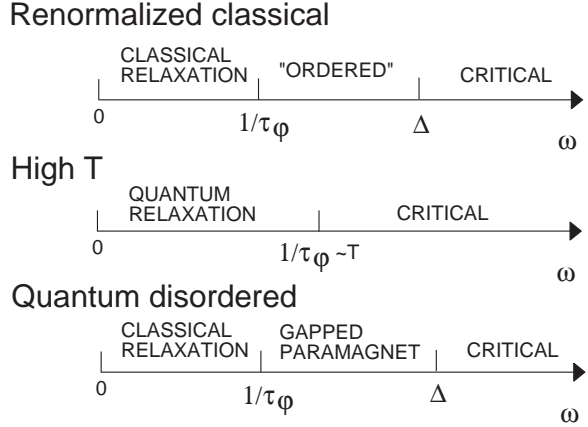}}
\caption{Crossovers as a function of frequency for the Ising model in the
different regimes of Fig~\protect\ref{fig3}. The correlations in the two classical relaxational
regimes are quite different from each other. The ``ordered'' regime is in quotes,
because there is no long-range order, and the system only appears ordered between
spatial scales $c/\Delta$ and $\xi_c$. In the renormalized-classical
and quantum-disordered regions $1/\tau_{\varphi} \sim T e^{|\Delta|/T}$.}
\label{fig8}
\end{figure}

The main features of the finite temperature physics of the quantum Ising model are
summarized in Figs~\ref{fig3} and Fig~\ref{fig8}. 
At short enough times or distances in
all three regions of Fig~\ref{fig3}, the systems displays critical fluctuations characterized by 
the dynamic susceptibility (\ref{ising34}). The regions are distinguished by their behaviors
at the low frequencies and momenta. In both the low $T$ regimes of Fig~\ref{fig3}
(renormalized classical and quantum disordered), the long time dynamics is
relaxational and is described by effective classical models.
The relaxation time, or equivalently, the phase coherence time, is of order
$(\hbar/k_B T) e^{(\mbox{energy gap})/k_B T}$, and is therefore much longer than
$\hbar / k_B T$. 
 In contrast, the
dynamics in the high
$T$ region is also relaxational, but involves quantum effects in an essential way, as was
described above. In this region the phase relaxation time is now of order $\hbar / k_B T$.

Finally a few remarks about experiments. While there are no experimental studies of the
Ising chain in a transverse field, the quantum relaxational dynamics of its
critical point $g=1$ is very closely related to the finite temperature dynamics
of $S=1/2$ Heisenberg spin chains. The spin correlators of the latter system
are given by expressions very similar to those in Section~\ref{sec:isingqc},
and the relaxational dynamics can has been experimentally probed in NMR experiments.
There is good agreement between theory and experiments, and the reader is referred
to some recent papers for further details~\cite{ssnmr,taki1,ssst}.

\section{$O(3)$ quantum rotors between one and three dimensions}
\label{chap:rotors2}

This section will examine ${\cal S}$ for $N=3$ in $d=1,2,3$:
the $N=3$ model describes the low energy physics of 
quantum antiferromagnets, and we will present a number of
experimental applications.

Many aspects of the physics
will be similar to those described in Section~\ref{chap:ising}
for the $N=1$, $d=1$ case.
However there are some genuinely new features which shall
be the focus of our attention. For $N\geq 2$, ${\cal S}$
has a continuous $O(N)$ symmetry, and associated conserved
charges, whose transport properties will be of interest to us.
Further, for $d>2$,
there is a
thermodynamic phase transition at a non-zero temperature: we shall briefly
consider
the subtle interplay between the
critical singularities of the non-zero temperature transition and those of 
the $T=0$ quantum critical point. 

As in the Ising case, it is useful to take a Hamiltonian point of view.
In this case (see Section~\ref{sec:largeN} for justification) the
associated model is the following Hamiltonian of $O(N)$ quantum rotors
on the sites $i$ of a
regular $d$ dimensional lattice (we will consider general $N\geq 2$ for
completeness):
\begin{equation}
H_R = \frac{J g}{2} \sum_i \stackrel{\leftrightarrow}{L^2}_i - J \sum_{\langle ij \rangle} 
\vec{n}_i \cdot
\vec{n}_j ,
\label{rotor0}
\end{equation}
where the sum $\langle ij \rangle$ is over nearest neighbors, $J>0$ is an overall energy scale,
and $g>0$ is a dimensionless coupling constant. The
$N$ component vectors  $\vec{n}_i$ are of unit length, $\vec{n}_i^2 = 1$,
and represent the orientation of the rotors on the surface of a sphere in
$N$-dimensional rotor space; as we will see later,
$\vec{n} \rightarrow \phi_{\alpha}$ in the mapping to the CQFT ${\cal S}$. 
The operators ${L}_{i\mu\nu}$ ($\mu < \nu$, 
$\mu$, $\nu = 1 \ldots N$) are the $N(N-1)/2$ components of the angular momentum 
$\stackrel{\leftrightarrow}{L}_i$ of the rotor: the first term in $H_R$ is the kinetic energy
of the rotor with $1/g$ the moment of inertia. The different components of $\vec{n}_i$
constitute a complete set of commuting observables and the state of the system
can be described by a wavefunction $\Psi (\vec{n}_i)$.
The action of $\stackrel{\leftrightarrow}{L}_i$ on $\Psi$ is given by the usual differential form 
of
the angular momentum
\begin{equation}
L_{i\mu\nu} = -i \left( n_{i\mu} \frac{\partial}{\partial n_{i\nu} }
- n_{i \nu} \frac{\partial}{\partial n_{i \mu} } \right).
\end{equation}
The commutation relations among the $\stackrel{\leftrightarrow}{L}_i$ and $\vec{n}_i$ can now
be easily deduced. We emphasize the difference of the rotors from Heisenberg-Dirac quantum spins:
the components of the latter at the same site do not commute, whereas the components of the
$\vec{n}_i$ do.

While there are no direct realizations of quantum rotors in nature, they are a useful
model of the low energy physics of a number of experimental systems. As we will see below,
a quantum rotor should be considered as a model for the physics of a {\em pair}
of antiferromagnetically coupled Heisenberg spins. 
The field $\vec{n}$ represents to the staggered antiferromagnetic order parameter,
while $\stackrel{\leftrightarrow}{L}$ is the total magnetization.
Thus quantum rotor models
describe the spin fluctuations of the spin-ladder compounds~\cite{azuma}, a rotor representing
the spins on each rung of the ladder. For technical reasons not explored
here, the rotor models also describe integer spin chains, and a large class
of Heisenberg antiferromagnets in higher dimensions. Finally, rotor models are
also useful in double-layer quantum Hall systems~\cite{dsz}, where superexchange
effects lead to an antiferromagnetic coupling between the electrons in the
two layers~\cite{lian}. These systems (excluding the double-layer
quantum Hall system for which the reader is referred to the literature)
will be described in more detail in the following subsections.

There is a strong analogy between the rotor Hamiltonian $H_R$ in (\ref{rotor0}) and the Ising
Hamiltonian $H_I$ in (\ref{ising1}). We will be looking at the transition between a magnetically
ordered state with $\langle \vec{n} \rangle \neq 0$ and $O(N)$ symmetry broken, and a quantum
paramagnet in which correlations of $\vec{n}$ are short ranged. As in the Ising model, it is the
exchage term, proportional to $J$, that favors the ordered state, while the `kinetic energy',
proportional to $Jg$ leads to fluctuations in the orientation of the order parameter and 
eventually
to loss of long-range order. The similarity between the two models will also be apparent in the
strong (large $g$) and weak coupling (small $g$) analyses in the following section.

\subsection{Limiting cases}
The pictures which emerge in the following two perturbative analyses are expected to hold on
either side of a quantum critical point at $g=g_c$.

\subsubsection{Strong coupling $g \gg 1$}
\label{sec:strongcoupling}
At $g=\infty$, the exchange term in $H_R$ can be neglected, and the Hamiltonian decouples into
independent sites, and can be diagonalized exactly. The eigenstates on each site
are the eigenstates of $\stackrel{\leftrightarrow}{L^2}$; for $N=3$ these are the states
\begin{equation}
\left| \ell, m \right\rangle_i~~~\ell = 0,1,2,\ldots,~ -\ell \leq m \leq \ell
\end{equation}
and have eigenenergy $J g \ell (\ell + 1)/2$. Compare this single site spectrum with that
of a pair of Heisenberg-Dirac spin $S$ quantum spins with an antiferromagnetic exchange $K$;
for a suitably chosen $K$ we get the same sequence of levels and energies but with a maximum
allowed value of $\ell = 2 S$. Assuming this upper cutoff is not crucial for the low energy
physics, we can use a single quantum rotor is an effective model for a {\em pair\/} of spins.

The ground state of $H_R$ in the large $g$ limit consists of the quantum paramagnetic state
with $\ell = 0$ on every site:
\begin{equation}
|0\rangle = \prod_i | \ell=0,m=0 \rangle_i
\end{equation}
Compare this with strong coupling ground state (\ref{ising1aa}) of the Ising model. Indeed, the
remainder of the strong coupling analysis of Section~\ref{sec:limiting} can borrowed here for the
rotor model, and we can therefore be quite brief. The lowest excited state is a `particle'
in which a single site has $\ell = 1$, and this excitation hops from site to site.
An important difference from the Ising model is that this particle is three-fold degenerate,
corresponding to the three allowed values $m$. The dynamic susceptibility has a quasiparticle 
pole
at the energy of this particle, and odd particle continua above the three particle threshold.

\subsubsection{Weak coupling, $g \ll 1$}
\label{sec:weakcoupling}
At $g=0$, the ground state breaks $O(N)$ symmetry, and all the $\vec{n}_i$ vectors orient 
themselves
in a common, but arbitrary direction. Excitations above this state consist of `spin waves' which 
can
have an arbitrarily low energy. This is a crucial difference from the Ising model, in which there
was an energy gap above the ground state. The presence of gapless spin excitations is a direct
consequence of the continuous $O(N)$ symmetry of $H_R$: we can make very slow deformations
in the orientation of $\langle \vec{n} \rangle$, and get an orthogonal state whose
energy is arbitrarily close to that of the ground state. Explicitly, for $N=3$, and a ground 
state
polarized along $(1,0,0)$ we parametrize
\begin{equation}
\vec{n} (x, t) = (1, \pi_1 (x, t), \pi_2 (x, t))
\end{equation}
where $|\pi_1|, |\pi_2| \ll 1$, and look at the linearized equations of motion for $\pi_1$, 
$\pi_2$.
A standard calculation then gives harmonic spin waves with energy $\omega = ck$
($c$ is the spin wave velocity); their wavefunctions can then be constructed using harmonic
oscillator states.

\subsection{Continuum theory and large $N$ limit}
\label{sec:largeN}

To obtain the path integral representation of the quantum mechanics of $H_R$,
we interpret the $\vec{n}_i$ as the co-ordinates of particles constrained to move on 
the surface of a sphere in $N$ dimensions: we can then simple use the standard
Feynman path integral representation of single-particle quantum mechanics.
After taking the continuum limit, such a procedure gives
\begin{displaymath}
Z = \int {\cal D} \vec{n} (x, \tau) \delta ( \vec{n}^2 ( x, \tau) - 1 )
\exp\left( - \int_0^{1/T} d \tau \int d^d x {\cal L} \right)
\end{displaymath}
\begin{equation}
{\cal L} = \frac{N}{2 c \tilde{g}} \left[  c^2 \left( \frac{\partial \vec{n} }{\partial x_i}
\right)^2 + 
\left( \frac{\partial \vec{n}}{\partial
\tau}
\right)^2 \right] 
\label{rotor1}
\end{equation}
Here $c \sim Ja$ is a velocity which will turn out to be the spin wave velocity,
and we have set
$\hbar = k_B = 1$. The prefactor of $N$ is for future convenience.
The coupling constant $\tilde{g} \sim
g a^{d-1}
$ has the dimensions of
$(\mbox{length})^{d-1}$; we will not use the original $g$ in $H_R$ further in this discussion,
and we will drop the tilde in $\tilde{g}$ from now. The
above action is valid only at long distances and times, so there is an implicit cutoff  above
momenta of order
$\Lambda \sim 1/a$ and frequencies of order
$c
\Lambda$. Our main interest here shall be the universal physics at scales much smaller than
$\Lambda$. It is now also apparent that if we convert the fixed length field $\vec{n}$
to the ``soft'' spin $\phi_{\alpha}$, the action in (\ref{rotor1}) maps to the
model ${\cal S}$.

A useful tool in the analysis of (\ref{rotor1})
is the limit of a
large number of components of $\vec{n}$ {\em i.e.\/} expansion in $1/N$~\cite{CSY}. 
For $N\geq 3$, 
the $N=\infty$ solution already contains the proper dilineation of all the crossovers
and phase transitions for all $d$. The description of the dynamic properties however requires
a fairly subtle analysis of the fluctuations. As all of these issues have been discussed
at length in the literature, here we will merely set-up the framework
of the $N=\infty$ theory, and then proceed to a physical discussion of the properties
of the various regimes for $d=1,2,3$.

The $N=\infty$ solution is quite easy to set up, at least in the
phase without long range order in the order parameter $\vec{n}$; we will
not discuss the ordered phase~\cite{CSY} here. We rescale the
$\vec{n}$ field to
\begin{equation}
\vec{\tilde{n}} =
\sqrt{N}
\vec{n},
\label{rotor1aa}
\end{equation} 
and impose the $\vec{\tilde{n}}^2 = N$ constraint by a Lagrange multiplier,
$\lambda$.
The action is then quadratic in the $\vec{\tilde{n}}$ field, which can then be
integrated out to yield
\begin{equation}
Z = \int {\cal D} \lambda \exp \left[ -
\frac{N}{2} \left( \mbox{Tr} \ln ( - c^2 \partial_i^2 -  \partial_{\tau}^2 + i
\lambda ) - \frac{i}{c g} \int_0^{1/T} d \tau \int d^d x  \lambda \right)\right]
\label{rotor1a}
\end{equation}
The action now has a prefactor of $N$, and the $N=\infty$ limit of the functional
integral is therefore given exactly by its saddle point value. We assume that the
saddle-point value of $\lambda$ is space and time independent, and given by
$i \lambda = \sigma^2$. The saddle-point equation determining the value of the parameter
$\sigma^2$ is 
\begin{equation}
\int^{\Lambda} \frac{d^d k}{(2 \pi)^d} T \sum_{\omega_n}
\frac{1}{c^2 k^2 + \omega_n^2  + \sigma^2} = \frac{1}{cg},
\label{rotor2}
\end{equation}
where the sum over $\omega_n$ extends over the Matsubara frequencies $\omega_n = 2 n
\pi T$, $n$ integer. It is also not difficult to see that
the retarded dynamic susceptibility $\chi (k, \omega)$, which is the two-point
correlator of the $n$ field, is given by
\begin{equation}
\chi (k, \omega) = \frac{c g/N}{c^2 k^2 - (\omega + i \delta)^2  + \sigma^2},
\label{rotor3}
\end{equation}
at $N=\infty$, with $\delta$ a positive infinitesimal.
The Eqns (\ref{rotor2},\ref{rotor3}) are the central results of the $N=\infty$
theory; in spite of their extremely simple structure, these equations contain a
great deal of information, and it takes a rather subtle and careful analysis to
extract the universal information contained in them~\cite{CSY}. 
We will not go into these technical details here, but will be satisfied by 
describing the different physical regimes predicted by 
(\ref{rotor2},\ref{rotor3}) and
other analyses of $H_R$.

\subsection{Dynamics in one dimension: application to spin-gap compounds}
\label{sec:diffusion}

Solution of (\ref{rotor2}) in $d=1$ shows that there is an energy gap, $\Delta$,
for all positive values of $g$. Even for very small values of $g$, where one might expect
long-range order, and 
the weak coupling solution of (\ref{sec:weakcoupling}) to apply, quantum fluctuations are
strong enough to always drive the system paramagnetic. For small $g$, the energy gap
behaves like
\begin{equation}
\Delta \sim c \Lambda e^{- 2 \pi /g}.
\label{valdeld1}
\end{equation}
This behavior can be confirmed by a more sophisticated renormalization 
group analysis~\cite{polyakov}. The implication of this important result is
that the large $g$ analysis of Section~\ref{sec:strongcoupling} actually
applies for {\em all\/} values of $g$ in $d=1$.

\begin{figure}[t]
\epsfxsize=3.3in
\centerline{\epsffile{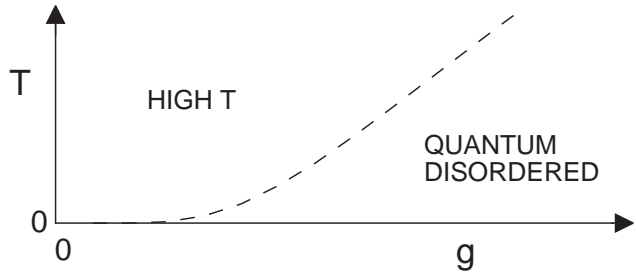}}
\caption{Crossovers of $H_R$ (Eq. (\protect\ref{rotor0})) in $d=1$
for $N\geq 3$. The phase boundary is at $\Delta\sim T$, with $\Delta$
given by (\protect\ref{valdeld1}).}
\label{fig9}
\end{figure}

The finite $T$ crossovers can also be deduced by a solution of (\ref{rotor2})~\cite{joli},
and are sketched in Fig~\ref{fig9}. 
The quantum critical point of Fig~\ref{fig3} has
effectively been moved to $g=0$, and there are now only $T$ distinct limiting
physical regimes: ({\em i\/}) $T
\ll \Delta$, where strong quantum fluctuations create a singlet paramagnetic
ground state, and we have a few thermally excited elementary 
excitations---for $N=3$ these are
a {\em triplet} of spin 1 particles with  energy $(\Delta^2 + c^2 k^2)^{1/2}$ 
at momentum $k$, and
correspond (crudely) to breaking a singlet valence bond between two 
neighboring spins in the 
underlying antiferromagnet; ({\em ii\/}) $\Delta
\ll T \ll c \Lambda$,
the high $T$ regime of the continuum theory, where quantum fluctuations are
marginally subdominant to thermal fluctuations~\cite{luscher} 
(by a factor of $1/\ln(T/\Delta)$), and we can locally describe
the system in terms of a {\em doublet\/} of left/right circularly polarized
spin-waves about an ordered state---however, interactions of thermally excited
spin waves lead again to a paramagnetic state.

Below we will present an {\em exact} theory for dynamic and transport
properties in the $T \ll \Delta$ regime~\cite{sd}. There is as yet no complete theory
for dynamics in the $T \gg \Delta$ regime, and its description
 remains an important open problem.

The basic approach followed for $T \ll \Delta$ is very similar to that taken
for the quantum-disordered region of the Ising chain in Section~\ref{sec:isingqd}.
The central difference here is that the quasi-particle excitations have an
additional spin label, $m= 1 \ldots N$. We will present our discussion below
for the physical case $N=3$, although the generalization to arbitrary $N$ is immediate.
This label on the quasiparticles is associated with the conserved $O(N)$
charge, which is the total spin $\sum_i \vec{L}_i$. We shall restrict our attention
here to the dynamics associated with the transport of this conserved charge:
this will be described by obtaining the exact long-time form of the correlator
\begin{equation}
C_{\alpha\beta} (x,t) = \left\langle L_{\alpha} (x,t) L_{\beta} (0,0) \right\rangle
\end{equation}
For the case $N=3$, the second-rank antisymmetric tensor indices $\mu\nu$ on $L$
have been replaced by a single vector index $\alpha$ or $\beta$.
The method described below can also be used to obtain dynamic correlators of the
$\vec{n}$ field: this will be discussed elsewhere~\cite{sd2}.

\begin{figure}[t]
\epsfxsize=2.7in
\centerline{\epsffile{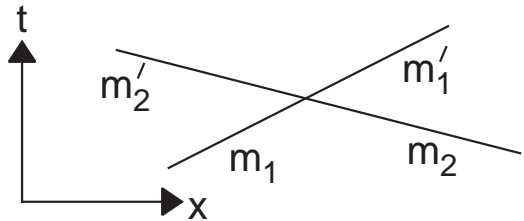}}
\caption{Two particle collision described by the $S$-matrix
(\protect\ref{smatrix}). The momenta before and after the collision are
the same, so the figure also represents the spacetime trajectories of
the particles. }
\label{fig10}
\end{figure}

There are two key observations that allow our exact computation for $T\ll
\Delta$. The first, as in the Ising chain, is that as the
density of particles $\sim e^{-\Delta/T}$, and their mean spacing is much
larger than their thermal de-Broglie wavelength $\sim c/(\Delta T)^{-1/2}$;
as a result the particles can be treated semiclassically. 
The density of particles with each spin $m$ ($m = -1, 0, 1$ for $N=3$)
is now given by the expression (\ref{eq:density}), and the total density
therefore equals $3 \rho$.
The second observation is that collisions between these particles are
described by their known two-particle $S$-matrix~\cite{zama}, and only a
simple limit of this $S$-matrix is needed in the low $T$ limit. The r.m.s.
velocity of a thermally excited particle $v_T = c (T/\Delta)^{1/2}$, and
hence its `rapidity' $\sim v_T / c \ll 1$. In this limit, the $S$-matrix
for the process in
Fig~\ref{fig10} is~\cite{zama}
\begin{equation}
 {S}^{m_1 m_2}_{m^{\prime}_1,
m^{\prime}_2} = (-1) \delta_{m_1 m^{\prime}_2} \delta_{m_2 m^{\prime}_1}.
\label{smatrix}
\end{equation}
In other words, the excitations behave like impenetrable
particles which preserve their spin in a collision. 
As in the Ising chain, energy and momentum conservation in $d=1$ require that these
particles simply exchange momenta across a collision (Fig~\ref{fig10}).
 The $(-1)$ factor in (\ref{smatrix}) can be interpreted as the phase-shift of
repulsive scattering between slowly moving bosons in $d=1$. Indeed, 
the simple form of
(\ref{smatrix}) is due to the slow motion of the particles, 
and is not a special feature of relativistic
continuum theory: it can be shown that (\ref{smatrix}) also holds for lattice
Heisenberg spin chains in the limit of vanishing velocities.

\begin{figure}[t]
\epsfxsize=3.3in
\centerline{\epsffile{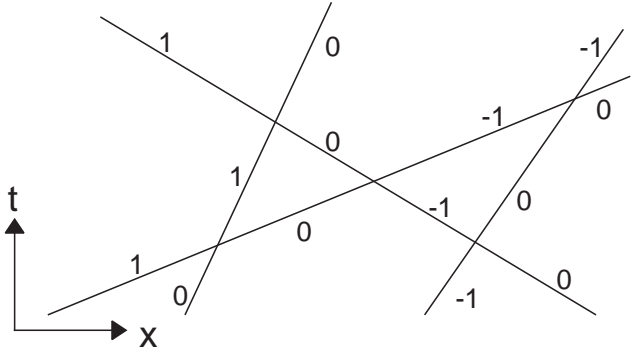}}
\caption{A typical set of particle trajectories contributing to $C(x,t)$.
Each trajectory represents paths moving both forward and backward in time,
and the $(-1)$ phase at each collision is neutralized by its time-reversed
contribution.
The particle co-ordinates are $x_k (t)$, with the labels $k$ chosen so that
$x_k(t) \leq x_l(t)$ for all $t$ and $k < \ell$.
Shown on the trajectories are the values of the particle spins $m_k$ which
are independent of $t$ in the low $T$ limit.}
\label{fig11}
\end{figure}

As in Sections~\ref{sec:isingrc} and~\ref{sec:isingqd}, we represent $C(x,t)$ as a `double time'
path integral, and in the classical
limit, stationary phase is achieved when the trajectories are
time-reversed pairs of classical paths (Fig~\ref{fig11}).
Each trajectory has a spin label, $m$, which obeys (\ref{smatrix}) at each collision.
The label, $m$, is assigned
randomly at some initial time with
equal probability,
but then evolves in time as discussed above (Fig~\ref{fig11}).
We label the particles consecutively from left to right by an integer
$k$; then their spins $m_k$ are independent of $t$, and we
denote their trajectories $x_k (t)$. The longitudinal correlation $C_{33}$ is then 
given by the
correlators of the classical observable
\begin{equation}
L_3 (x, t) = \sum_k m_k \delta( x - x_k (t))
\label{m3class}
\end{equation}
in the classical ensemble defined in Section~\ref{sec:isingrc} and above. 
Now because the spin and
spatial co-ordinates are independently distributed, the correlators
of $m_k$ and $x_k$ factorize. The $m_k$ are uncorrelated, and therefore
$\langle m_k m_{\ell} \rangle  = (2/3) \delta_{k \ell}$
We therefore have
\begin{equation}
C_{33} (x - x', t - t') = \frac{2}{3} \sum_k \left\langle \delta(x - x_k (t))
\delta(x' - x_k(t'))
\right\rangle
\label{ab}
\end{equation}
This result involves the self two-point correlation of a given
particle $k$, which follows a complicated trajectory in the way we
have labeled the particles ({\em e.g.\/} the trajectory of the $-1$ in Fig~\ref{fig11}).
Fortunately, precisely this correlator was considered
three decades ago by Jepsen~\cite{jepsen} and a little later by
others~\cite{lebowitz}; they showed that, at sufficiently long times, this
correlator has a Brownian motion form.
Inserting their results into (\ref{ab}), we obtain the results presented below.

An important property of the results is that they can written in a
`reduced' scaling form~\cite{CSY}, much like that found in Sections~\ref{sec:isingrc}
and~\ref{sec:isingqd}.
The characteristic spatial length ($L_x$) and time ($L_t$) can
 be chosen to be
\begin{equation}
L_x = \frac{1}{3\rho} ~~~~~L_t = \frac{1}{3\rho} \left( \frac{\Delta}{2
c^2 T} \right)^{1/2}.
\label{defl}
\end{equation}
Notice $L_x \sim c e^{\Delta/T}/\sqrt{\Delta T}$ is the mean spacing between the particles ,
and $L_t \sim e^{\Delta/T}/T$ is a
typical time between particle collisions.
Our final result is
\begin{equation}
C_{\alpha\beta} (x,t) =  6 \rho^2 F \left( \frac{|x|}{L_x} , \frac{|t|}{L_t}
\right) \delta_{\alpha\beta}
\end{equation}
where $F$ is a universal scaling function. A complete expression for $F$ is
given in Ref~\cite{sd}. Here we note the important limiting forms.
For short times $F$ has the ballistic form
\begin{equation}
F ( \bar{x}, \bar{t} ) \approx e^{-\bar{x}^2/\bar{t}^2}/\bar{t} \sqrt{\pi},
\end{equation}
which is the auto-correlator of a classical ideal gas in $d=1$, and holds for
$|\bar{t}| \ll |\bar{x}| \ll 1$. In contrast, for 
$|\bar{t}| \gg 1, |\bar{x}|$ it crosses over to the {\em diffusive \/} form
\begin{equation}
F (\bar{x} , \bar{t}) \approx \frac{e^{-\sqrt{\pi}\bar{x}^2/2 \bar{t}}}{(4 \pi
\bar{t}^2)^{1/4}}
\label{larget}
\end{equation}
In the original dimensionful units, (\ref{defl}) and (\ref{larget})
imply a spin diffusion constant, $D_s$, given exactly by
\begin{equation}
D_s = \frac{c^2 e^{\Delta/T}}{ 3 \Delta } .
\label{diffres}
\end{equation}
To the best of our knowledge, this is first exact result for the spin diffusivity
in any paramagnetic spin system.

The above long-time diffusive form of the magnetization correlations have
important consequences for NMR experiments on one-dimensional compounds with a
spin gap. The comparison with experiments requires generalization of the above
theory to include an external magnetic field $H$. This has been discussed elsewhere~\cite{sd}.
From this computation one finds that the longitudinal nuclear spin relaxation rate, $1/T_1$,
is given in the weak field limit $H \ll \Delta, L_t^{-1}$ by
\begin{displaymath}
\frac{1}{T_1} = \frac{\Gamma T \chi}{\sqrt{2 D_s H}} =
\frac{\Gamma \Delta e^{-3 \Delta / 2 T}}{c^2} \sqrt{\frac{3 T}{\pi
H}} ~~;~~\chi_u = \frac{e^{-\Delta / T}}{c} \sqrt{ \frac{2 \Delta}{ \pi T}},
\end{displaymath}
where $\Gamma$ is a nuclear hyperfine coupling, 
and $\chi_u$ is the paramagnetic spin susceptibility (in the rotor model, this
is the linear response to a field $H$ under which $H_R \rightarrow
H_R - H \sum_i L_{i3}$). For experimental
comparisons, a striking property of the above, pointed out to us by M.~Takigawa, is that the low
$T$ activation gaps for $1/T_1$ ($\Delta_{1/T_1}$) and $\chi$
($\Delta_{\chi}$) satisfy $\Delta_{1/T_1}/\Delta_{\chi} = 3/2$.

\begin{figure}[t]
\epsfxsize=3.5in
\centerline{\epsffile{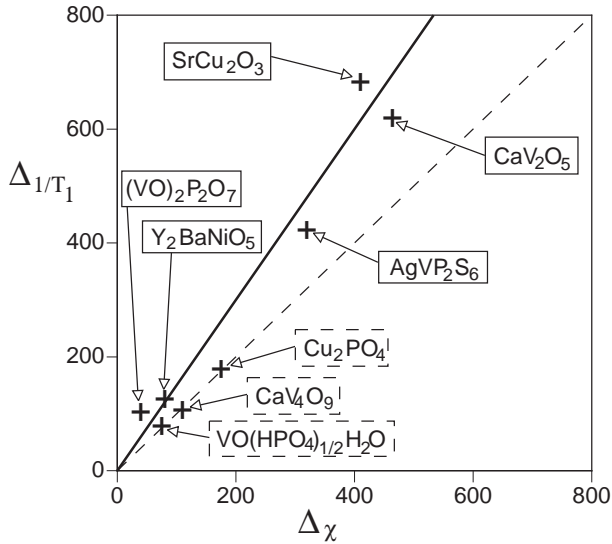}}
\caption{Measured values of the activation gaps $\Delta_{1/T_1}$ and $\Delta_{\chi}$
for a number of spin gap systems. The one-dimensional compounds are enclosed by
full rectangles, while others have dashed rectangles (these are either two-dimensional
or consist of isolated dimers). The full line is the theoretically predicted ratio
of $3/2$ and is in good agreement with the trends in the $d=1$ systems.
The other systems have a ratio close to unity, and this is expected to be the
case when elastic impurity-scattering of the quasiparticles dominates.}
\label{fig12}
\end{figure}

Values for the activation
gaps $\Delta_{1/T_1}$ and $\Delta_{\chi}$ have been quoted for a number of
experimental systems~\cite{taki,shimizu,azuma}, and it has consistently
been found that $\Delta_{1/T_1}$ is larger than $\Delta_{\chi}$. 
The above theoretical results provide a natural explanation for this
puzzling experimental trend. This is further explored in Fig~\ref{fig12},
which was kindly provided to us by H.~Yasuoka.

We have also compared~\cite{sd2} the detailed $H$ and $T$ dependence
of the $1/T_1$ measurements on Ref~\cite{taki} with the predictions of our theory.
These measurements have enough dynamic range to explore the ballistic-diffusive 
crossover described above, and the results are in good agreement with 
theoretical predictions.

\subsection{Two dimensions}
\label{sec:rotor2d}
 
Solution of the $N=\infty$ equation (\ref{rotor2}) in $d=2$ yields a phase diagram 
essentially identical to that shown in Fig~\ref{fig3} for the $d=1$ quantum Ising model.
A renormalization group analysis~\cite{CHN} shows that these large $N$ results
are qualitatively correct for the physical case of $N=3$.
A detailed description of the physical properties in the different regions of the
phase diagram of Fig~\ref{fig3} for the $d=2$ $O(3)$ rotor model has been obtained 
in the $1/N$ expansion~\cite{CSY}. We will not describe this analysis here as
the physical properties are quite similar to those for the $d=1$ Ising
model, which we have discussed in some detail in Section~\ref{chap:ising}.
Further, the results of the large $N$ theory have been reviewed 
elsewhere~\cite{statphys,madrid}.
 
We do note one important difference
from the $d=1$ Ising model, however. The ordered phase ($g < g_c$) of the $d=2$ $O(3)$ rotor
model has gapless spin-wave excitations above the ground state, unlike the Ising model
which had a gap. Consequently, we need a different energy scale to characterize
the deviation of the ground state from the quantum critical point $g=g_c$. 
A convenient choice is the spin stiffness, $\rho_s$, which has the dimensions of energy
in $d=2$. The non-zero $T$ properties of the rotor model will now be universal functions
of the dimensionless ratio $\rho_s / T$ for $g< g_c$. 
The renormalized-classical region is therefore present for $T < \rho_s$,
while the ``High T'' region is $T > \rho_s$.
There is an energy gap, $\Delta$, towards a triplet of quasiparticle excitations in the 
quantum-disordered phase present for $g>g_c$, and the nonzero $T$
crossovers in this case are very
similar to those of the $d=1$ Ising model: the physics is ``quantum-disordered''
for $T < \Delta$, and ``High T'' for $T > \Delta$.

The $d=2$ $O(3)$ rotor model is expected to describe the low energy
properties two dimensional quantum antiferromagnets whose low energy states
are well described by fluctuations of a {\em collinear} antiferromagnetic spin
ordering. The $S=1/2$ spin-1/2 square lattice antiferromagnet, realized in
${\rm La}_2{\rm CuO}_4$ and related insulators, is an important example of such a system,
and we now discuss the relationship of its properties to those of the rotor model.  
By a comparison with low temperature behavior of the correlation length observed
in neutron scattering
measurements, Chakravarty {\em et al.}~\cite{CHN}
convincingly argued that
the ground state of this model has long-range N\'{e}el order, {\em i.e.} it maps onto
the rotor model with a coupling $g<g_c$. 
Their analysis was restricted to the renormalized classical region of Fig~\ref{fig3}
{\em i.e.} $T \ll \rho_s$, where static properties reduce to those of an effective
classical rotor model. A theory for the dynamic properties in this region
was presented in Ref~\cite{tyc} by postulating classical equations of motion
for the classical rotor model.

Next, we turn to a question~\cite{cs}
which arises naturally from the structure of Fig~\ref{fig3}: does the
$S=1/2$ square lattice antiferromagnet
crossover into the ``High T'' regime at temperatures $T > \rho_s$ ? 
The existence of such a crossover is subject to the condition that it
occur at temperatures small enough so that the mapping to rotor model
remains valid. 
Strong evidence for the ``High T'' regime was found in Ref~\cite{cs}
by a comparison of the $1/N$ expansion of the rotor model
with numerical and experimental measurements of the uniform susceptibitlity.
The magnitude and temperature dependence of the $1/T_1$~\cite{imai1}
and $1/T_{2G}$~\cite{imai2} nuclear magnetic relaxation rate measurements
on ${\rm La}_2{\rm CuO}_4$ were also found to be in good agreement
with theory~\cite{CSY,css}. Convincing evidence for the
renormalized-classical to ``High T'' crossover was presented 
by Sokol {\em et al.}~\cite{rajiv} and Elstner {\em et al.}~\cite{elstner}
using a high temperature series expansion for a number of static
correlators of the $S=1/2$ square
lattice antiferromagnet. (Elstner {\em et al.}~\cite{elstner}
also examined the $S=1$ square lattice antiferromagnet,
and found no evidence for the ``High T'' regime; this 
indicates, as expected, that the effective value
of $g$ for $S=1$ is significantly smaller than $g_c$.)
It was also argued early on~\cite{cs} that
the $T$ dependence of the correlation length, which was central to 
the conclusions of Chakravarty {\em et al.} , would not exhibit a
clear signature of the ``High T'' regime. This point was
subsequently reiterated by Greven {\em et al.}~\cite{greven}.

A more complete understanding of the physics of the ``High T''
regime should emerge from a careful study of the 
wavevector, frequency, and temperature dependence of the
imaginary part of the staggered dynamic spin susceptiblity. This should obey
a scaling form~\cite{sy} much like that obtained earlier for the
Ising model in (\ref{ising35}):
\begin{equation}
\chi^{\prime\prime} (k, \omega ) = 
\frac{{\cal A}}{T^{2-\eta}} \Phi_{\chi} \left ( \frac{\hbar c k}{k_B T},
\frac{\hbar \omega}{k_B T} \right),
\end{equation}
where ${\cal A}$ is a non-universal amplitude and $\eta$ is the
anomalous dimension of $\vec{n}$ field at the quantum-critical point
($\eta \approx 0.03$ for $d=2$, $N=3$).
Detailed theoretical results for the form of the universal scaling
function $\Phi_{\chi}$ have been obtained~\cite{sy,CSY} in the
$1/N$ expansion. Unfortunately there are no existing results
 for $\chi'' (k, \omega )$ in ${\rm
La}_2{\rm CuO}_4$ at $T > \rho_s$, as the neutron scattering intensity
becomes quite small at higher temperatures. However there are other
two-dimensional square lattice antiferromagnets with a smaller $\rho_s$~\cite{mcm}
which are good candidates for such a study.

This is a good point to 
make some speculative remarks about the {\em doped}
antiferromagnet ${\rm La}_{2-x}{\rm Sr}_{x}{\rm CuO}_4$.
The $1/T_1$ measurements of Ref~\cite{imai1} have the striking property
that while the low $T$ values of $1/T_1$ are quite strongly doping
dependent, at higher $T$ $1/T_1$ approaches a value which is both
temperature and doping-concentration independent. This appears to receive
a natural explanation within the framework of a rotor model
quantum-critical point. Assuming the main effect of doping is merely to 
shift the value of the effective coupling constant $g$, the properties
of the ``High T'' region of Fig~\ref{fig3} are determined primarily by
the value of $T$ alone, and are insensitive to the precise value of $g$:
indeed, in this region, the system does not `know' whether $g$ is smaller
or larger than $g_c$. Further, the $T$ dependence is predicted to
be $1/T_1 \sim T^{\eta}$~\cite{cs,CSY}, and given the small value of $\eta$,
this is essentially $T$-independent. Lastly, the coefficient of $T^{\eta}$
has also been estimated~\cite{CSY} and is consistent with the experimental
value.

We turn now to other antiferromagnets in $d=2$ which have a collinear
N\'{e}el ordered ground state. An interesting system in which it is possible
to tune the value of the effective coupling $g$ across $g_c$
is the double layer square lattice $S=1/2$ antiferromagnet~\cite{double}. 
This model consists of spin-1/2 Heisenberg spins on two adjacent square
lattices, with an intralayer antiferromagnetic exchange $J$ and an interlayer
antiferromagnetic exchange
$K$. The ratio
$K/J$ acts much like the dimensionless coupling $g$, with the large $K/J$ phase a gapped quantum
paramagnet of singlet pairs of spins in opposite layers, and the small $K/J$ phase N\'{e}el
ordered. Numerical simulations have been carried out on this model by Sandvik and
collaborators~\cite{sandvik}, and the critical point $g=g_c$ identified rather reasonably well. 
A number of universal amplitude ratios have been studied at this point, and all
results are now in good agreement with the $1/N$ expansion on the $O(N)$ quantum rotor model.
High temperature~\cite{elstner,awsss} and strong coupling~\cite{gelfand}
series expansions on the double-layer model also reach a similar conclusion.

An exciting recent development has been the remarkably high precision study of 
quantum-critical properties in the 1/5-depleted square lattice by
Troyer {\em et al.}~\cite{troyer2}. This is another model in which $g$ can be
tuned across $g_c$ by varying the ratio of exchange constants. This particular
lattice has the additional feature that the mapping to the rotor model 
is expected to be incomplete: there are additional Berry phases generated
by the dynamics of the Heisenberg spins undergoing a ``hedgehog'' tunneling event~\cite{cavo}.
It was argued~\cite{CSY} 
that such perturbations should be ``dangerously irrelevant'', and that
they do not modify any of the universal scaling predictions of the rotor model.
The results of Troyer {\em et al.} are in excellent agreement with the
critical properties of the rotor model~\cite{troyer2}, thus supporting the
neglect of the Berry phases.

\subsection{Three dimensions}

\begin{figure}[t]
\epsfxsize=8.3in
\centerline{\epsffile{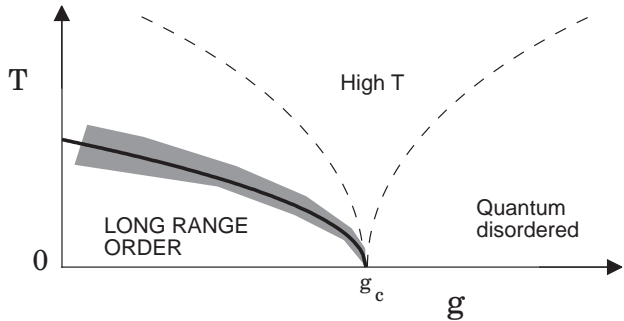}}
\caption{Phase diagram for the $O(N)$ rotor model with for $N>2$ and $d=3$. A similar
diagram applies for the cases with $N=1,2$ and $d=2,3$. The dashed lines are crossovers,
while the full line is the locus of finite temperature phase transitions. The shaded region
is where the reduced classical scaling functions apply.}
\label{fig13}
\end{figure}

Solution of the $N=\infty$ equation (\ref{rotor2}) now yields the phase digram shown
in Fig~\ref{fig13}~\cite{madrid}.
The new feature is presence of a finite temperature phase transition: this lies entirely
within the low $T$ region for $g< g_c$ and is not a direct property of the quantum critical 
point. The well-studied classical critical behavior appears only within the
shaded region of Fig~\ref{fig13}. The quantum-critical scaling functions contain
this classical criticality as ``reduced scaling functions'' in much the same way
that in the $d=1$ Ising model the classical function (\ref{apy2}) was contained in
the more general quantum function in (\ref{ising27}).
This phenomenon can be studied with the $1/N$ expansion~\cite{madrid},
and has also recently been described in an expansion in $\epsilon = 3-d$~\cite{epsilon}.
Also, because $d=3$ is the upper-critical dimension of ${\cal S}$, there are logarithmic
violations of scaling near the quantum-critical point.

Normand and Rice~\cite{normand} have proposed 
an interesting recent experimental realization of such a three-dimensional quantum critical
point in ${\rm LaCuO}_{2.5}$. This is a spin-ladder compound in which the ladders are
moderately strongly coupled in three dimensions. By varying the ratio of the intra-ladder
to inter-ladder exchange it is possible to drive such an antiferromagnet
across a $d=3$ quantum critical point separating N\'{e}el ordered and quantum paramagnetic
phases. The uniform susceptibility has a $T^2$ dependence at intermediate $T$,
which is characteristic of the ``High T'' region in $d=3$. The entire $T$ dependence
of the uniform susceptibility has been compared with large $N$ and quantum Monte Carlo
simulations of the quantum-critical model~\cite{troyer2}, with good agreement.

\section{Quantum relaxational transport in two dimensions}

Transport properties of the model ${\cal S}$ were considered earlier
in Section~\ref{sec:diffusion}. There we determined the spin diffusion
constant in the quantum-disordered regime ($T \ll \Delta$)
of the $N=3$ model in $d=1$. Here, following Ref~\cite{sd1}, 
we shall describe transport
for $N \geq 2$
in $d=2$ in the ``High T'' region (Fig~\ref{fig1} and~\ref{fig3}),
 where the
dynamics is quantum-relaxational (Fig~\ref{fig8}) . 
The methods discussed here can also be applied to the
quantum-disordered region of the $N \geq 2$, $d=2$ models, but we shall
not present that extension here.  Let us also recall that there is one
region for which there is no quantitative theory of transport
phenomena: the ``High $T$''
region for $N \geq 3$ in $d=1$; this region should be 
experimentally accessible in $S=2$ spin chains.

We shall present our discussion using the language of the $N=2$
system, although closely related results apply to all $N \geq 2$.
The $N=2$ model describes a superfluid-Mott insulator transition
in a lattice boson model with short-range repulsive interactions~\cite{fwgf}.
The complex superfluid order parameter $\Psi$ is related to the field
$\phi_{\alpha}$ by 
\begin{equation}
\Psi = (\phi_1 + i \phi_2)/\sqrt{2}
\end{equation} 
(and similarly to $\vec{n}$).
It therefore serves a starting point for understanding superfluid-insulator
transitions in disordered thin films~\cite{hebard,lg}
and Josephson junction arrays~\cite{jj} and the transitions among quantum Hall 
states~\cite{shahar,sankar,klz}.

Rather than describing the transport in terms of the diffusivity,
it is more convenient here, both theoretically and experimentally,
to characterize it using the conductivity, $\sigma$; the two 
quantities are, of course, connected by the Einstein relation. 
We consider the response of ${\cal S}$ to a spatially
varying ``chemical potential'', $\mu$, (this is the field that couples to the 
conserved charge for $N=2$; for the $N=3$ case we called it the
``magnetic field'') under which the rotor model Hamiltonian transforms as
\begin{equation}
H_R \rightarrow H_R - Q \sum_i \mu(i) L_i.
\end{equation}
Here $Q$ is the charge of the $\Psi$ field. Note 
that for $N=2$ there is only one generator $L$ which therefore does not carry
any vector indices. In an open system, the expectation value of the current
(this is the current associated with the conserved charge) will be proportional
to the gradient of $\mu$, and the proportionality constant
is $\sigma$. We will be interested here in the $\omega$ and $T$ dependence
of $\sigma$ at the critical point $g=g_c$, as that controls the value of
$\sigma$ in the ``High T'' region. We begin by writing down the expected
scaling form this must satisfy. For this we need the scaling and engineering
dimensions of $\sigma$. Rather general arguments~\cite{fgg,cha,conserve,shahar}
show that in $d=2$, $\sigma$ has scaling dimension zero, and engineering dimensions
of the quantum unit of conductance $Q^2 / \hbar$. This leads to the scaling form
\begin{equation}
\sigma = \frac{Q^2}{\hbar} \Sigma\left( \frac{\hbar \omega}{k_B T}
\right)
\label{condscal}
\end{equation}
where $\Sigma$ is a universal scaling function whose properties we will 
describe here. 

It is clear from (\ref{condscal}) that the experimentally
measured d.c. conductivity is determined by the universal number $\Sigma (0)$.
In contrast, any computation of the conductivity performed at $T=0$,
in which the $\omega \rightarrow 0$ limit is taken subsequently will determine
the number $\Sigma ( \infty )$. It was argued quite generally 
by Cha {\em et al.}~\cite{cha}
and Wallin {\em et al.}~\cite{wallin} that $\Sigma$ is in independent of
$\omega/T$, which therefore also implies that $\Sigma (0) = \Sigma (\infty)$.
As a result, there were a number of analytic 
computations~\cite{fgg,cha,mpaf,wz,kz,fz,ih2,ww,cfw} and an exact diagonalization 
study~\cite{runge} of the
value of $\Sigma ( \infty )$ in a variety of models which display a quantum-critical
point in $d=2$. Quantum Monte Carlo studies~\cite{cha,wallin,swgy,mtu,bls} 
determined the conductivity
by analytic continuation from imaginary time data at $\omega = 2 \pi n T i$,
with $n \geq 1$ integer; as the smallest value of $\omega$ is $2 \pi T i$,
these studies were effectively also in the regime $|\omega| \gg T$. 

\begin{figure}
\epsfxsize=3.2in
\centerline{\epsffile{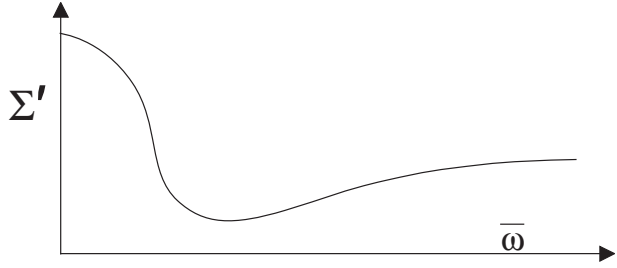}}
\caption{A sketch of the expected form of the real part, $\Sigma'$,
of the universal scaling function
$\Sigma$ appearing in the scaling form (\protect\ref{condscal})
for the conductivity, as a function of $\overline{\omega} = \hbar \omega / k_B T$.
There is a Drude-like peak from the inelastic scatterings between thermally excited
carriers at $\overline{\omega}$ of order unity. At larger $\overline{\omega}$, there is 
a crossover to the collisionless regime where
$\Sigma' \rightarrow \Sigma (\infty)$ .
}
\label{fig14}
\end{figure}

The picture of quantum relaxational dynamics we have reviewed in this paper
makes it quite clear that the underlying assumption of these works
is incorrect: $\Sigma$ is in fact, {\em not} independent of $\omega / T$.
As we have discussed here, the characteristic property of the ``High T''
region is that there is a phase relaxation time $\tau_{\varphi} \sim
\hbar / k_B T$. Dynamic order parameter fluctuations also carry charge, and therefore 
inelastic collisions between thermally excited charge-carrying
excitations will lead 
to a transport relaxation time $\tau_{\rm tr}$. As the typical
energy exchanged in a collision is $k_B T$, $\tau_{\rm tr}$ is
 also of order $\tau_{\varphi}$,
and therefore
\begin{equation}
\frac{1}{\tau_{\rm tr}} \sim \frac{k_B T}{\hbar}.
\label{tautr}
\end{equation}
The missing coefficient in (\ref{tautr}) is a universal number whose value will
depend upon the precise definition of $1/\tau_{\rm tr}$. It is perhaps worth noting
explicitly here that (\ref{tautr}) holds for all values of the dynamic 
exponent $z$. The relationship (\ref{tautr}) could be violated if interactions
were dangerously irrelevant at the quantum critical point, a possibility
we shall not discuss here.
Now general considerations~\cite{kb} suggest that there are two qualitatively different
regimes of charge transport at non-zero frequencies
\newline
(I) $\omega \tau_{\rm tr} \ll 1$ ($\hbar \omega \ll k_B T$): the hydrodynamic,
collision-dominated, incoherent regime,
where charge transport is controlled by repeated, inelastic scatterings between
pre-existing thermally, excited carriers; the conductivity should exhibit
a `Drude' peak as a function of frequency.\newline
(II) $\omega \tau_{\rm tr} \gg 1$ ($\hbar \omega \gg k_B T$): 
the high frequency, collisionless, phase-coherent regime, 
where the excitations created by the external perturbation are solely 
responsible for transport, and collisions with thermally excited carriers can be neglected.
\newline
These physical arguments lead to a suggested form for the function $\Sigma ( \overline{\omega})$
shown in Fig~\ref{fig14}.
We have assumed that $\Sigma (0) > \Sigma ( \infty )$, and this will be the 
case in the specific calculation for ${\cal S}$ discussed below.
The minimum at $\omega \sim k_B T/\hbar$ also appears in the calculation
below, but it is not clear whether that is a general feature.

\begin{figure}
\epsfxsize=3in
\centerline{\epsffile{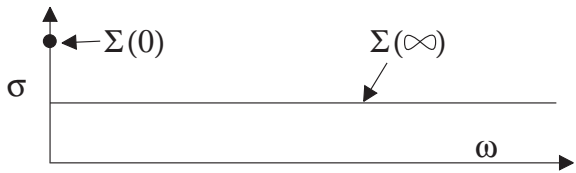}}
\caption{Universal form of the conductivity
$\sigma ( \omega , T \rightarrow 0 )$
in $d=2$; the vertical scale is measured in units of $\hbar/Q^2$. 
Only the $\omega =0$ value is given by the universal number 
$\Sigma ( 0 )$. For all $\omega > 0$, $(\hbar/Q^2) \sigma = \Sigma ( \infty )$.
}
\label{fig15}
\end{figure}

It is also interesting to consider the implication of Fig~\ref{fig14} for
$\sigma ( \omega, T \rightarrow 0)$.
This is illustrated in Fig~\ref{fig15} which plots the form of
$(\hbar/Q^2) \sigma ( \omega, T \rightarrow 0)$ in $d=2$: its value at $\omega =0$ is
given by $\Sigma (0)$, while for all $\omega > 0$ it equals $\Sigma ( \infty )$.
Note the difference from Fermi liquid theory, where the Drude peak becomes
a delta function with non-zero weight as $T \rightarrow 0$. In the present
case,
the weight in the Drude-like peak vanishes like $\sim T$ as $T \rightarrow 0$,
and reduces to the {\em single} point $\omega = 0$ where the conductivity is
given by $\Sigma ( 0)$. 

We note that the above discussion
implies results that are rather remarkable from a traditional
quantum transport point of view. There have been a number of previous situations
in which charge transport properties have been found to be universally related
to the quantum unit of conductance, $e^2 / h$; these include the 
quantized Landauer conductance of ballistic transport in one-dimensional
wires, and the universal conductance fluctuations of mesoscopic metals~\cite{ucf}. 
However in all
previous cases, these universal properties have arisen in a ``phase-coherent'' regime,
{\em i.e.\/} they are associated with physics at scales shorter than the mean
distance between inelastic scattering events between the carriers. For the
case of a $d=2$ quantum critical point discussed above, the 
universal number $\Sigma ( \infty )$
is associated with quantum coherent transport, and is therefore 
the analog of these earlier results. In contrast, the value of $\Sigma (0)$ is controlled by
repeated inelastic scattering events, and therefore the d.c. transport
is clearly in what would traditionally
be identified as the ``incoherent'' regime. Nevertheless, we have argued above that
$\Sigma (0) $ is a universal number, and the d.c. conductance remains
universally related to $(Q=e)^2 / h$.

Let us now turn to the computation of $\Sigma$ for the specific model ${\cal S}$.
The $T>0$ transport properties of ${\cal S}$ in $d=2$ were considered
briefly by Cha {\em et al.}~\cite{cha}. They concluded that the continuum model ${\cal S}$
had $\sigma = \infty$ for all $T>0$, and that additional lattice umklapp scattering effects
were needed to degrade the current carried by thermally excited carriers, 
and to yield a $\sigma < \infty$. However, in Section~\ref{sec:diffusion} we a found
finite diffusivity $D_s$ for $N=3$, $d=1$ at $T>0$, and we claim that a similar
result holds here for $N=2$, $d=2$. The basic point can be made quite simply by a
glance at Fig~\ref{fig11}. Note that in this $d=1$ model, the transport of {\em momentum}
is indeed ballistic, and is represented by the straight line trajectories which extend
for all time. However, when we consider the transport of {\em charge}, we have to follow
the motion of the $N$-valued $O(N)$ index carried by each trajectory: this motion 
is seen in Fig~\ref{fig11} to be rather complicated, and 
was shown in Section~\ref{sec:diffusion} to be diffusive. There is thus an 
essential difference between
momentum and charge transport when the excitations carry a $O(N)$ flavor index,
and it appears to have been overlooked in Ref~\cite{cha}.

We will outline the recent computation of $\Sigma$ in an expansion in
$\epsilon = 3-d$~\cite{sd1}. For small $\epsilon$ the quartic interaction, $u_0$,
in ${\cal S}$ reaches a universal (to leading order in $\epsilon$) fixed point
 value~\cite{bgz}
\begin{equation}
u_0 = \frac{24 \pi^2}{5} \epsilon
\end{equation}
in the vicinity of the quantum critical point.
So the scattering of the elementary excitations, and the non-ballistic motion
of the charge current, only occurs with an amplitude of order $\epsilon$. 
Such a weak-coupling $S$-matrix should be compared with the complementary
strong-coupling
$S$-matrix in (\ref{smatrix}) for $d=1$: in the latter case we had a maximal
scattering situation as momentum and spin labels interchanged their 
mutual pairing in every collision. Also in contrast to the real-space
formulation of transport taken for $d=1$ in Section~\ref{sec:diffusion},
the weak-coupling situation here is best analyzed in momentum space.
We wish to work with same charge eigenstates we considered in 
Section~\ref{sec:diffusion}, and so we express the field $\Psi$
in the associated normal modes 
\begin{equation}
\Psi (\bx,t) = \int \frac{d^d k}{(2 \pi)^d}
\frac{1}{\sqrt{2\varepsilon_k}}
\left( a_{+} (\bk, t) e^{i \bk \cdot \bx} + a_{-}^{\dagger} (\bk, t) e^{- i
\bk \cdot \bx} \right),
\end{equation}
where $\varepsilon_k = (k^2 + m^2)^{1/2}$, with $m$ a renormalized $T$-dependent
mass~\cite{sd1}.
The operators $a_{\pm}$ annihilate particles with  $O(2)$
charges $\pm1$ (in Section~\ref{sec:diffusion}, where we considered $O(3)$,
there were three such particles). Next, transport theory requires 
the momentum space distribution functions
\begin{equation}
f_{\lambda} ( \bk , t) = \left\langle a^{\dagger}_{\lambda} ( \bk, t)
a_{\lambda} ( \bk , t) \right \rangle,
\end{equation}
with $\lambda = \pm 1$,
in terms of which a quantum Boltzmann equation for transport in a weakly-scattering
system can be obtained~\cite{kb}. We will not display the explicit form of this
equation, or discuss its solution here: the reader should consult Ref~\cite{sd1}.
That analysis leads to an explicit solution for $\Sigma$ which has the qualitative
form of Fig~\ref{fig14}. We quote quantitative results for the two limiting values~\cite{sd1,fz}:
\begin{eqnarray}
\Sigma ( 0) &=& \frac{0.1650}{\epsilon^2} + \ldots \nonumber \\
\Sigma' ( \overline{\omega} \rightarrow \infty ) &=& 
\frac{2^{1-2d} \pi^{1-d/2}}{d \Gamma (d/2)} \left(
1 + {\cal O} (\epsilon^2 ) 
\right) \overline{\omega}^{1-\epsilon} 
\label{i1}
\end{eqnarray}
to leading order in $\epsilon$. Notice the singular $\epsilon$ dependence of the first
result: this is signal of a ``boundary layer'' in $\Sigma$ of width $\omega \sim \epsilon^2 T$
which constitutes the ``Drude peak'' of Fig~\ref{fig14}. This boundary layer explicitly 
demonstrates a claim made in Section~\ref{intro}: the operations of expansion
in $\epsilon$ 
and of analytic continuation in $\omega$ do not commute.
The structure of the higher-order corrections
to the first result in (\ref{i1}) is quite complex, and was generally discussed in 
Ref~\cite{epsilon}.
For the superfluid insulator transition in $d=2$ with $Q=2e$,
 (\ref{i1}) implies a universal d.c. conductivity 
\begin{equation}
\sigma (0) = 
2 \pi \Sigma (0) \frac{4e^2}{h} \approx 1.037 \frac{4e^2}{h}
\label{i1b}
\end{equation} 
This result is remarkably close to the value, $4 e^2 /h$ (argued~\cite{mpaf,wz} to be
related to self-duality), and to the results of many experiments~\cite{lg}.
Indeed, it was conjectured in Ref~\cite{sd1} that $2 \pi \Sigma (0) = 1$ exactly
for $N=2$ in $d=2$. 
Definitively establishing this self-duality however
requires techniques other than expansion in $\epsilon = 3-d$, or $1/N$,
as it is only possible precisely at $d=2$, $N=2$.
Finding such techniques is an important challenge for future work.

\section{Conclusion}
We have reviewed the universal, non-zero temperature dynamic properties of the
CQFT ${\cal S}$ for a number of representative cases, and discussed
their experimental implications. Among the cases omitted, there are some
that show phenomena qualitatively different from those discussed here: 
\newline
({\em i}) The case $N=2$, $d=1$ has a line of critical points and a Kosterlitz-Thouless
end point at $T=0$. Its paramagnetic phase is similar to the $N=3$, $d=1$
model discussed in Section~\ref{sec:diffusion}.
\newline
({\em ii}) For $d>3$ there is violation of hyperscaling, which leads to explicit
dependence of finite $T$ behavior on the bare value of $u_0$ and qualitatively
different crossovers~\cite{epsilon}.

\acknowledgements
I thank Profs. A.~Skjeltorp and D.~Sherrington for the kind invitation
to lecture at the NATO Geilo school, and all the participants for stimulating
interactions. I am also grateful to my collaborators on the topics
reviewed here: A.~Chubukov, K.~Damle, T.~Senthil, R.~Shankar,
A.~Sokol, J.~Ye and A.P.~Young.
This research was supported by the U.S. National Science Foundation 
under Grant DMR-96-23181.

\end{document}